\documentclass[a4paper,11pt]{article}
\pdfoutput=1
\usepackage{jheppub} 
\usepackage[T1]{fontenc}
\usepackage{rotating}
\usepackage{pgf}

\newcommand*\PD[1]{\doswap#1\relax}
\def\doswap#1,#2,#3\relax{\left( \frac{\partial #1}{\partial #2}\right)_{#3}}

\newcommand{\beq}{\begin{equation}}
\newcommand{\eeq}{\end{equation}}

\newcommand{\ba}{\begin{eqnarray}}
\newcommand{\ea}{\end{eqnarray}}

\begin{document}
\title{Gravitational collapse of thin shells: Time evolution of the holographic entanglement entropy}

\rightline{BI-TP 2014/28}
\rightline{HIP-2015-01/TH}
\rightline{TUW-15-03}
\rightline{NIKHEF-002}

\author[a]{Ville Ker\"{a}nen,}
\author[b]{Hiromichi Nishimura,}
\author[c]{Stefan Stricker,}
\author[d]{Olli Taanila,}
\author[e]{Aleksi Vuorinen}

\affiliation[a]{Rudolf Peierls Centre for Theoretical Physics, University of Oxford, 1 Keble Road,\\ Oxford OX1 3NP, United Kingdom}
\affiliation[b]{Faculty of Physics, Bielefeld University,
D-33615 Bielefeld, Germany}
\affiliation[c]{Institute of Theoretical Physics, Technical University of Vienna, Wiedner Hauptstr.~8-10,\\ A-1040 Vienna, Austria}
\affiliation[d]{Nikhef, Science Park 105, 1098 XG Amsterdam, The Netherlands}
\affiliation[e]{Department of Physics and Helsinki Institute of Physics, P.O.~Box 64,\\ FI-00014 University of Helsinki, Finland}

\emailAdd{vkeranen1@gmail.com}
\emailAdd{nishimura@physik.uni-bielefeld.de}
\emailAdd{stricker@hep.itp.tuwien.ac.at}
\emailAdd{olli.taanila@iki.fi}
\emailAdd{aleksi.vuorinen@helsinki.fi}

\abstract{We study the dynamics of gravitationally collapsing massive shells in AdS spacetime, and show in detail how one can determine extremal surfaces traversing them. The results are used to solve the time evolution of the holographic entanglement entropy in a strongly coupled dual conformal gauge theory, which is is seen to exhibit a regime of linear growth independent of the shape of the boundary entangling region and the equation of state of the shell. Our exact results are finally compared to those of two commonly used approximation schemes, the Vaidya metric and the quasistatic limit, whose respective regions of validity we quantitatively determine.}

\maketitle

\section{Introduction}

By now, holography has established itself as one of the main tools used to gain insights into the out-of-equilibrium dynamics of strongly coupled field theories. Mapping the process of thermalization into black hole formation in asymptotically Anti-de Sitter (AdS) spacetime, gauge/gravity \cite{Maldacena:1997re,Gubser:1998bc,Witten:1998qj} methods have already solved several outstanding problems motivated by both heavy ion and condensed matter physics that have long eluded solutions using traditional field theory techniques (for reviews, see e.g.~\cite{Gubser:2009md,CasalderreySolana:2011us,DeWolfe:2013cua,Brambilla:2014jmp}). This can be largely attributed to the absence of competition: Perturbative methods typically fail already at moderate couplings, while time-dependent quantum phenomena are outside the realm of lattice Monte-Carlo simulations.

Important recent advances in applied holography include a fully dynamical description of shock wave collisions in strongly coupled ${\mathcal N}=4$ Super Yang-Mills (SYM) theory \cite{Chesler:2010bi, Wu:2011yd, Casalderrey-Solana:2013aba, vanderSchee:2012qj,vanderSchee:2013pia, Chesler:2015wra} as well as extensive work on the evolution of entropy-like quantities such as the holographic entanglement entropy (HEE) \cite{AbajoArrastia:2010yt,Albash:2010mv,Baron:2012fv,Liu:2013iza,Liu:2013qca,Abajo-Arrastia:2014fma,Balasubramanian:2011ur,Pedraza:2014moa, Auzzi:2013pca, Alishahiha:2014cwa}. At the same time, technical leaps have been taken in the incorporation of inhomogeneities and anisotropies in thermalization dynamics \cite{Chesler:2008hg,Heller:2012km,Chesler:2012zk,Heller:2013oxa,Balasubramanian:2013rva}, the development of a formalism to evaluate out-of-equilibrium Green's functions \cite{CaronHuot:2011dr,Mukhopadhyay:2012hv,Balasubramanian:2012tu,Keranen:2014lna}, as well as the first studies of thermalization dynamics away from the infinite coupling limit 
\cite{Steineder:2012si,Steineder:2013ana,Stricker:2013lma,Baron:2013cya} and in non-conformal backgrounds \cite{Craps:2013iaa}.

The above list of references clearly reflects an ongoing pursuit to take the holographic description of equilibration dynamics closer to the physical systems realized in nature, which are typically characterized by complicated initial states, finite coupling strength and $N_c$, as well as broken conformal invariance. In this approach, one is typically confined to determining rather simple observables such as the temporal and spatial evolution of energy density or pressure. A different line of research concentrates on the simplest thermalization models available, but attempts to compute more complicated quantities, such as various off-equilibrium Green's functions and other non-local observables. One prominent example of such models involves the gravitational collapse of an infinitesimally thin but massive shell in AdS space \cite{Danielsson:1999zt,Danielsson:1999fa}; following these papers, several works have addressed a variety of physical phenomena including particle production rates \cite{Baier:2012ax, Baier:2012tc}, the chiral magnetic effect \cite{Lin:2013sga}, jet quenching \cite{Caceres:2012px} and even elliptic flow \cite{Muller:2013ila}. Most of these calculations, however, apply the so-called quasistatic approximation and assume the time scale related to the collapse to be parametrically larger than the other scales of interest, thus effectively considering the shell a static object \cite{Lin:2008rw}.

In a preceding paper \cite{Keranen:2014zoa}, we reported results from a set of calculations inspecting the falling shell model in a fully dynamical setup, where the shell follows a physical trajectory solved from the Einstein equations. The quantities considered in this context were the HEE and the Causal Holographic Information (CHI), which are both examples of geometric probes whose determination reduces to finding the area of some bulk hypersurface. As this involved rather complicated calculations requiring finding and matching extremal surfaces and geodesics in a time-dependent background, one of the aims of our current paper is to walk the reader through the technical details of this work. In addition, we will, however, present a considerably more thorough analysis of the HEE, comparing in particular its time evolution to results obtained in the quasistatic approximation and in the Vaidya metric. Here, we will find that during all times at least one of these approximation schemes is in a good quantitative agreement with the full results. We will also analyze the dynamics of the collapsing shell itself, and provide the full details of the construction of a coordinate system continuous at the location of the shell, briefly introduced already in \cite{Keranen:2014zoa}.

In references \cite{Liu:2013iza,Liu:2013qca}, it was noticed that in the Vaidya spacetime the entanglement entropy of large boundary regions exhibits linear increase in time for an extended period. The coefficient of this increase, $v_\text{E}$, quantifies the rate, at which the time evolution entangles the subsystem to its surroundings. The authors of \cite{Liu:2013iza,Liu:2013qca} proposed an interesting conjecture that the value of $v_\text{E}$ computed for a collapse from AdS to the AdS-Schwarzschild spacetime might provide an upper bound for the rate of entanglement production in any relativistic quantum field theory. Furthermore, it was argued here that the rate $v_\text{E}$ is a property of the final equilibrium state only, as it is only affected by the metric of the final black hole. One way of testing this proposal is to consider different initial states that evolve towards the same thermal state at late times --- an exercise straightforwardly implementable in the collapsing shell model. As we will see, in all of our results the rate $v_\text{E}$ is indeed seen to be independent of the details of the shell trajectory, i.e.~of the way the non-equilibrium initial state is prepared. Thus, we find evidence supporting the picture that $v_\text{E}$ is a property of the final equilibrium state only.

Our paper is organized as follows: First, in section 2 and the corresponding appendices A, B and C, we provide technical details of our calculations, including solving for the shell dynamics, constructing a coordinate system that is continuous across the shell, and deriving continuity conditions for geodesics and extremal surfaces at the shell. In section 3, we then analyze the solutions to the shell equation of motion (EoM), while section 4 as well as appendix D are devoted to deriving the HEE and analyzing the corresponding results. Finally, in section 5 we compare our numerical findings to the quasistatic and Vaidya limits, analyzing the regions of validity of these approximation schemes, and in section 6 we draw our conclusions.

\section{Details of the calculation}

In this section, we introduce the machinery needed to obtain the time evolution of the HEE we are after. To this end, we first introduce our collapsing shell setup and derive the EoM of a shell falling in AdS$_5$ spacetime in section \ref{sec:dynamics}. Then, we derive a coordinate system continuous at the shell in section \ref{sec:junctionconditions}, which we use to write down junction conditions for extremal surfaces and more generic geometric probes intersecting the shell. Several details of the calculations are left to appendices A--C.

\subsection{Setup and shell dynamics}

\label{sec:dynamics}

Just as in \cite{Keranen:2014zoa}, we work in a spacetime characterized by a negative cosmological constant, into which we immerse a thin massive shell, whose energy momentum tensor is proportional to a delta function in the radial coordinate.\footnote{Similarities of the thin shell setup and the fully back-reacted numerical solution of the Einstein-Klein-Gordon system are discussed in \cite{Garfinkle:2011tc}.} Since both inside and outside the shell, the space is a solution to vacuum Einstein equations, we choose the inside metric to be that of an empty AdS Poincar\'{e} patch and the outside metric the AdS-Schwarzschild solution with Schwarzschild radius $r_h$,
\begin{eqnarray}
\label{eq:metric} ds^2 &=& -f_\pm(r)\,dt^2 + \frac{dr^2}{f_\pm(r)} + r^2 d\mathbf{x}^2\,, \\
\label{eq:fpm}
f_\pm(r) &=& \begin{cases} \hfill r^2 - \frac{r_h^4}{r^2}, \hfill & \text{ if $r > r_s$} \\ \hfill r^2, \hfill & \text{ if $r<r_s$}\end{cases} \, .
\end{eqnarray}
Here we have introduced a notation that we will be using throughout the calculation, where the subscripts $+$ and $-$ refer to quantities evalauted outside and inside the shell, respectively. It is important to note that although the metric functions $f_\pm$ themselves are time independent, the location of the shell $r_s$, and thus the location of the discontinuity, are time dependent.  The radial coordinate $r$ and the spatial coordinates $\mathbf{x}$ are in addition assumed to be continuous at the shell. This means that there are two different and \emph{a priori} unrelated time coordinates $t_+$ and $t_-$, which we will later relate to each other.

The coordinates on the shell are chosen to be the proper time of the shell and the spatial coordinates $\mathbf{x}$, denoted by
\begin{equation}
\left[ \xi^i  \right] = \left( \tau, \mathbf{x} \right) \, .
\end{equation}
The embedding of the shell in the five-dimensional space is then given by
\begin{equation}
\left[ y^\mu \right] = \left( t_{s\pm}(\tau),r_s(\tau),\mathbf{x}\right) \, ,
\end{equation}
where $\mu$ is an index running over the five coordinates of the $\mathrm{AdS}_5$ space. Requiring that $\tau$ is the proper time of the shell, we can further relate $t_{s\pm}$ and $r_s$ to each other by writing
\begin{equation}
ds^2 = -d\tau^2 = -f_\pm \,\dot{t}_{s\pm}^2\,d\tau^2 + \frac{\dot{r}_s^2}{f_\pm}\,d\tau^2 \, .
\end{equation}
Thus, the derivatives of $t_{s\pm}$ and $r_s$ with respect to the proper time of the shell --- denoted here by dots --- are related by
\begin{equation}
f_\pm \,\dot{t}_{s\pm} = \sqrt{f_\pm + \dot{r}_s^2}\;.
\end{equation}

In appendix \ref{sec:eom}, we derive the EoM of the shell, given by eq.~(\ref{eq:Israeli}). To evaluate its right-hand side, we need to specify the energy momentum content of the shell in the appropriate coordinate system. To this end, we employ the perfect fluid form,
\begin{equation}
S^{ij} = (\rho + p ) u^i u^j +p\,\gamma^{ij}, \label{eq:fluidem}
\end{equation}
where $u^i$ the four-velocity of the fluid and $\gamma^{ij}$ the induced metric on the shell. This is in fact the most general possible energy momentum tensor when imposing translational and rotational symmetry in the $\textbf{x}$ directions.  Since the time coordinate in the $\xi$ coordinate system is the proper time of the shell, the coordinate system is in the rest frame of the fluid, and thus $u = (1,\mathbf{0})$. The two independent non-zero components of equation (\ref{eq:Israeli}) then become
\begin{align}
\label{eq:betas}-\frac{3}{r_s} \left( \sqrt{f_-+\dot{r}_s^2} -\sqrt{f_+ +\dot{r}_s^2} \right) & = -8\pi g_5\,\rho\, ,\\
\frac{1}{\dot{r}_s} \frac{d}{d\tau}\left[r_s^2\left(  \sqrt{f_-+\dot{r}_s^2}-\sqrt{f_++\dot{r}_s^2}\right)\right] & = -8\pi g_5\,p\,r_s^2 \, ,
\end{align}
from which we can derive a scaling law for the energy density,
\begin{equation}
\frac{d}{dr_s} \left( \rho \,r_s^3\right) = -3\,r_s^2\,p \, .
\end{equation}
Considering the simple equation of state (EoS) $p = c\,\rho$, we finally obtain
\begin{equation}
\rho \propto r_s^{-3(1+c)} \, ,
\end{equation}
so that we can define a constant of motion $M$ satisfying
\begin{equation}
\label{eq:rhoscale} \frac{8}{3}\pi g_5 \, \rho = \frac{M}{r_s^{3(1+c)}}\, .
\end{equation}
When the shell is pressureless ($c=0$), the constant $M$ is directly related to the conserved mass of the shell.

After inserting eq.~(\ref{eq:rhoscale}) to (\ref{eq:betas}), we get as the final EoM of the shell
\begin{equation}
\dot{r}_s^2 = \frac{M^2}{4\,r_s^{4+6c}}-\frac{f_-+f_+}{2}+\frac{(f_--f_+)^2r_s^{4+6c}}{4\,M^2}\, ,
\end{equation}
where $f_-$ and $f_+$ are evaluated at the shell, $r=r_s$. It is noteworthy that the functional forms of $f_\pm$ are at this point still arbitrary, and that this equation is first order in time derivatives. The latter fact implies that solving it requires only one initial condition, e.g.~the value of $r_s$ at some known time $\tau$, while the initial velocity is encoded in the constant $M$. This equation can be interpreted as the non-linear generalization of the conservation of kinetic and potential energy in Newtonian mechanics.

If we now insert the explicit forms of $f_\pm$ from equation (\ref{eq:fpm}), we obtain from the above
\begin{equation}
\dot{r}_s^2 =- r_s^2 +\frac{r_h^4}{2\,r_s^2}+\frac{M^2}{4\,r_s^{4+6c}} + \frac{r_h^8 r_s^{6c}}{4\,M^2} \, . \label{shelleom1}
\end{equation}
Using as the initial conditions $r_s(\tau=0) = r_0$, $\dot{r}_s(\tau=0) = 0$, this allows us to solve the value of $M$ as
\begin{equation}
M^2 = r_0^{4+6c}\left[ \sqrt{f_+(r_0)}-\sqrt{f_-(r_0)}\right]^2 \, ,
\end{equation}
or using the explicit form of $f_\pm$,
\begin{equation}
M^2 = 2\,r_0^{6(1+c)} \left( 1- \frac{r_h^4}{2\,r_0^4} -\sqrt{1-\frac{r_h^4}{r_0^4}}\right) \, .
\end{equation}
Together with the equation of motion (\ref{shelleom1}), this determines how the shell falls as a function of its proper time. If one on the other hand wants to EoM of the shell in terms of the coordinate time, or possibly relate the discontinuous time coordinates on the two sides of the shell to each other, one has to further use the relation
\begin{equation}
\frac{dt_-}{dt_+} = \frac{\dot{t}_{s-}}{\dot{t}_{s+}} = \frac{f_+}{f_-}\sqrt{\frac{f_-+\dot{r}_s^2}{f_+ +\dot{r}_s^2}} \, ,
\end{equation}
which applies at the shell.

\subsection{The junction conditions}

\label{sec:junctionconditions}

In order to eventually determine the time evolution of the entanglement entropy in the boundary field theory, we must be able to solve minimal surfaces in the spacetime containing a moving shell. In particular, we need to know how to join the minimal surfaces across the shell, i.e.~how they refract at the shell. As we will review in section \ref{sec:entropy}, the determination of a minimal surface can be phrased as a variational problem, where one extremizes a functional of the generic form
\beq
S=\int d^n \sigma\,\mathcal{L}\left[x^{\mu}(\sigma),\partial_{a}x^{\mu}(\sigma),g_{\mu\nu}\right],\label{eq:genericfunctional}
\eeq
where $ \partial_a x^{\mu}(\sigma)=\partial x^{\mu}(\sigma)/\partial\sigma^{a}$,
with $\sigma$ denoting some set of coordinates on the minimal surface and $x^{\mu}(\sigma)$ encoding the embedding of the surface in the spacetime. In this section, we will work out the refraction conditions following from extremizing a generic functional of the form  (\ref{eq:genericfunctional}). Thus, the results we obtain can be applied to any geometric probes in the spacetime, such as geodesics, string worldsheets and minimal area surfaces.

Varying the action of eq.~(\ref{eq:genericfunctional}) leads to equations of motion for $x^{\mu}(\sigma)$, the Euler-Lagrange equations, that involve first derivatives of the metric. As we are dealing with a metric that is discontinuous, these equations will have delta function contributions from the derivatives. One way to derive junction conditions for $x^{\mu}(\sigma)$ would be to integrate the  EoMs across these singularities; in our case, this is, however, difficult to apply in practice, so we will use a different method. Namely, we will in the following explicitly construct a coordinate system, where the metric is continuous at the position of the shell. Working within it, the EoMs will have no delta function singularities, and therefore the solution $x^{\mu}(\sigma)$ and all its first derivatives $\partial x^{\mu}(\sigma)/\partial \sigma^a$ will be continuous across the shell. Then, to obtain the junction conditions in the original coordinate system, we simply perform a coordinate transformation back to the original coordinates, where the discontinuities in the derivatives reappear from discontinuities in the coordinate transformation.

To explicitly construct the coordinate system described above, we choose the timelike coordinate to be the proper time of the shell, $\tau$. Correspondingly, the required spatial coordinate is chosen to be the proper physical distance from the shell normal to it, which we denote by $\lambda$ and use to define our time slicing. Thus, our coordinate transformation has the form
\beq
(t_{\pm},r,\mathbf{x})\rightarrow (\tau,\lambda,\mathbf{x})\,,
\eeq
where a complication, however, arises from the fact that the normal vector of the shell is only defined at its location. This implies that we need to parallel transport this vector to cover the other parts of the spacetime. Intuitively, we start from the shell and then head out in the direction of the normal vector, parallel transporting it according to
\begin{equation}
\nabla_n n=0 \; .
\end{equation}
This requirement is clearly nothing but the geodesic equation, meaning that our new spatial coordinate is simply the physical distance from the shell along a spacelike geodesic normal to the shell at its location.

In order to determine the metric in this continuous coordinate system as well as to obtain the desired junction conditions, we need to know how the coordinates in the different coordinate systems are related to each other. Instead of obtaining explicit expressions for the coordinate transformation, it is, however, sufficient to merely calculate the values of the partial derivatives\footnote{Here we have introduced the notation $\PD{a,b,c}$ familiar from thermodynamics to keep in mind which parameter is held constant as the other one is varied. To make our expressions somewhat more compact, we have also suppressed the arguments of our functions: When using the coordinates $\tau$ and $\lambda$, $r$ and $t$ are functions of both of these variables, whereas the time and position of the shell, $t_s$ and $r_s$ are functions of  $\tau$ only. Furthermore, $f$ is a function of $r$ and thus of both $\tau$ and $\lambda$ while $f_s = f(r_s(\tau))$.}
\begin{equation}
\PD{t,\tau,\lambda} , \; \PD{t,\lambda,\tau} , \; \PD{r,\tau,\lambda} \; \text{and} \;\; \PD{r,\lambda,\tau}
\end{equation}
at the shell. This exercise is performed in appendix \ref{sec:cont}.

We will now proceed to compute the first total derivatives $d x^{\mu}/d \sigma^a$ in the outside patch, transform them to the new coordinate system, and then transform them further to the inside patch. Using the chain rule, we can write the necessary derivatives in the form
\begin{align}
\frac{dt_+}{d\sigma^a} &= \frac{d}{d\sigma^a} t_+\left(\tau(\sigma),\lambda(\sigma)\right) = \PD{t_+,\tau,\lambda}\frac{d\tau}{d\sigma^a} + \PD{t_+,\lambda,\tau}\frac{d\lambda}{d\sigma^a}\,,\\
\frac{dr_+}{d\sigma^a} &= \frac{d}{d\sigma^a} r_+\left(\tau(\sigma),\lambda(\sigma)\right) = \PD{r_+,\tau,\lambda}\frac{d\tau}{d\sigma^a} + \PD{r_+,\lambda,\tau}\frac{d\lambda}{d\sigma^a}\,,
\end{align}
where we will now drop the index $a$ from $\sigma^a$ to simplify our notation. From these expressions, we then solve
\begin{align}
\label{eq:dlambdadx}\frac{d\lambda}{d\sigma} & = \frac{\PD{r_+,\tau,\lambda}\frac{dt_+}{d\sigma}-\PD{t_+,\tau,\lambda}\frac{dr_+}{d\sigma}}{\PD{r_+,\tau,\lambda}\PD{t_+,\lambda,\tau}-\PD{r_+,\lambda,\tau}\PD{t_+,\tau,\lambda}}\,, \\
\label{eq:dsdx}\frac{d\tau}{d\sigma} & = \frac{\PD{r_+,\lambda,\tau}\frac{dt_+}{d\sigma}-\PD{t_+,\lambda,\tau}\frac{dr_+}{d\sigma}}{\PD{r_+,\lambda,\tau}\PD{t_+,\tau,\lambda}-\PD{r_+,\tau,\lambda}\PD{t_+,\lambda,\tau}} \, , 
\end{align}
which, when evaluated at the shell using the partial derivatives calculated in appendix \ref{sec:cont}, gives further
\begin{align}
\label{eq:dlambdadx+} \frac{d\lambda}{d\sigma} &= -\dot{r}_s \frac{dt_+}{d\sigma} + \frac{\sqrt{f_{s+}+\dot{r}_s^2}}{f_{s+}} \frac{dr_+}{d\sigma}\,, \\
\label{eq:dsdx+}  \frac{d \tau}{d\sigma} & = \sqrt{f_{s+}+\dot{r}_s^2}\frac{dt_+}{d\sigma}-\frac{\dot{r}_s}{f_{s+}}\frac{dr_+}{d\sigma} \; .
\end{align}

Next, we use the chain rule to express $dt/d\sigma$ and $dr/d\sigma$ in the inside patch,
\begin{align}
\frac{dt_-}{d\sigma} &= \PD{t_-,\tau,\lambda}\frac{d\tau}{d\sigma} + \PD{t_-,\lambda,\tau}\frac{d\lambda}{d\sigma}\,,\\
\frac{dr_-}{d\sigma} &= \PD{r_-,\tau,\lambda}\frac{d\tau}{d\sigma} + \PD{r_-,\lambda,\tau}\frac{d\lambda}{d\sigma}\,,
\end{align}
which, evaluated again at the shell, produces
\begin{align}
\frac{dt_-}{d\sigma} &= \frac{\sqrt{f_{s-} + \dot{r}_s^2}}{f_{s-}}\frac{d\tau}{d\sigma} + \frac{\dot{r}_s}{f_{s-}}\frac{d\lambda}{d\sigma}\,,\\
\frac{dr_-}{d\sigma} &= \dot{r}_s\frac{d\tau}{d\sigma} + \sqrt{f_{s-}+\dot{r}_s^2}\frac{d\lambda}{d\sigma} \, .
\end{align}
Finally, we insert eqs.~(\ref{eq:dlambdadx+}) and (\ref{eq:dsdx+}) into the above equations to get the junction conditions
\begin{align}
\label{eq:match1} \left.\frac{dt_-}{d\sigma}\right|_{r=r_s} &= \left.\frac{dt_+}{d\sigma}\right|_{r=r_s} \frac{\beta_{s-}\beta_{s+}-\dot{r}_s^2}{f_-} + \left.\frac{dr_+}{d\sigma}\right|_{r=r_s} \frac{\dot{r}_s}{f_- f_+}\left(\beta_{s+}-\beta_{s-}\right)\,,\\
\label{eq:match2} \left.\frac{dr_-}{d\sigma}\right|_{r=r_s}  &= \left.\frac{dt_+}{d\sigma}\right|_{r=r_s} \dot{r}_s\left( \beta_{s+}-\beta_{s-}\right) + \left.\frac{dr_+}{d\sigma}\right|_{r=r_s} \frac{1}{f_+}\left( \beta_{s+}\beta_{s-}-\dot{r}_s^2\right)\, ,
\end{align}
where 
$\beta_{s\pm} = \sqrt{f_{s\pm}+\dot{r}_s^2}$. As a consistency check, we verify that in the limit where the shell vanishes, $f_- \to f_+$, both of these relations become identities. Also, in the limit where the velocity of the shell approaches the speed of light $\dot{r}_s\rightarrow\infty$, the junction conditions reduce to the ones previously found in the Vaidya spacetime, cf.~e.g.~\cite{Liu:2013qca}.

Interestingly, the above matching conditions are valid in a space with an arbitrary dimensionality, and one only needs to modify the metric functions  $f_+$ and $f_-$ in eq.~(\ref{eq:fpm}). Also, in Appendix \ref{genjun} we show how the conditions get modified, if one takes as the starting point of the calculation a more generic metric, where the $dt^2$ and $dr^2$ components are a priori not related to each other.

\section{Properties of the shell motion \label{sec:shellmot}}

In this section, we perform a systematic study of the solutions of the shell EoM for different values of the EoS parameter $c$, defined through $p=c\, \rho$. For brevity, we will here denote $t_{s\pm} (\tau)$ by simply $t_\pm$.

\subsection{Simple example: $c=-1/3$}
Let us start by considering in detail the case of $c=-1/3$, which exhibits the same qualitative features as the more general cases studied later, but is computationally somewhat simpler. For this value of $c$, the equations of motion namely reduce to
\begin{align}
\dot{r}_s^2&=-r_s^2+r_0^2\Big(\frac{r_0}{r_s}\Big)^2\,,\label{eq:rs}
\\
\dot{t}_+&=\frac{r_s\sqrt{r_0^4-r_h^4}}{r_s^4-r_h^4}\,,\label{eq:tp}
\end{align}
of which we can solve the first one by direct integration, producing
\begin{equation}
\tau=\frac{1}{r_0^2}\int_{r_s}^{r_0}\frac{dr r}{\sqrt{1-\Big(\frac{r}{r_0}\Big)^4}}=\frac{1}{2}\arccos\Big(\frac{r_s}{r_0}\Big)^2 \, ,
\end{equation}
or equivalently
\begin{equation}
r_s(\tau)=r_0\sqrt{\cos(2\tau)}\,.
\end{equation}
From here, we see that for small and negative $\tau$ the shell  heads towards the boundary, while  at $r=r_0$ or $\tau=0$ it turns around and collapses. At the proper time $\tau=\pi/4$, the shell reaches the singularity at $r=0$. 

\begin{figure}[t]
\begin{center}
\includegraphics[scale=0.55,trim={0cm 0cm 4cm 19cm},clip]{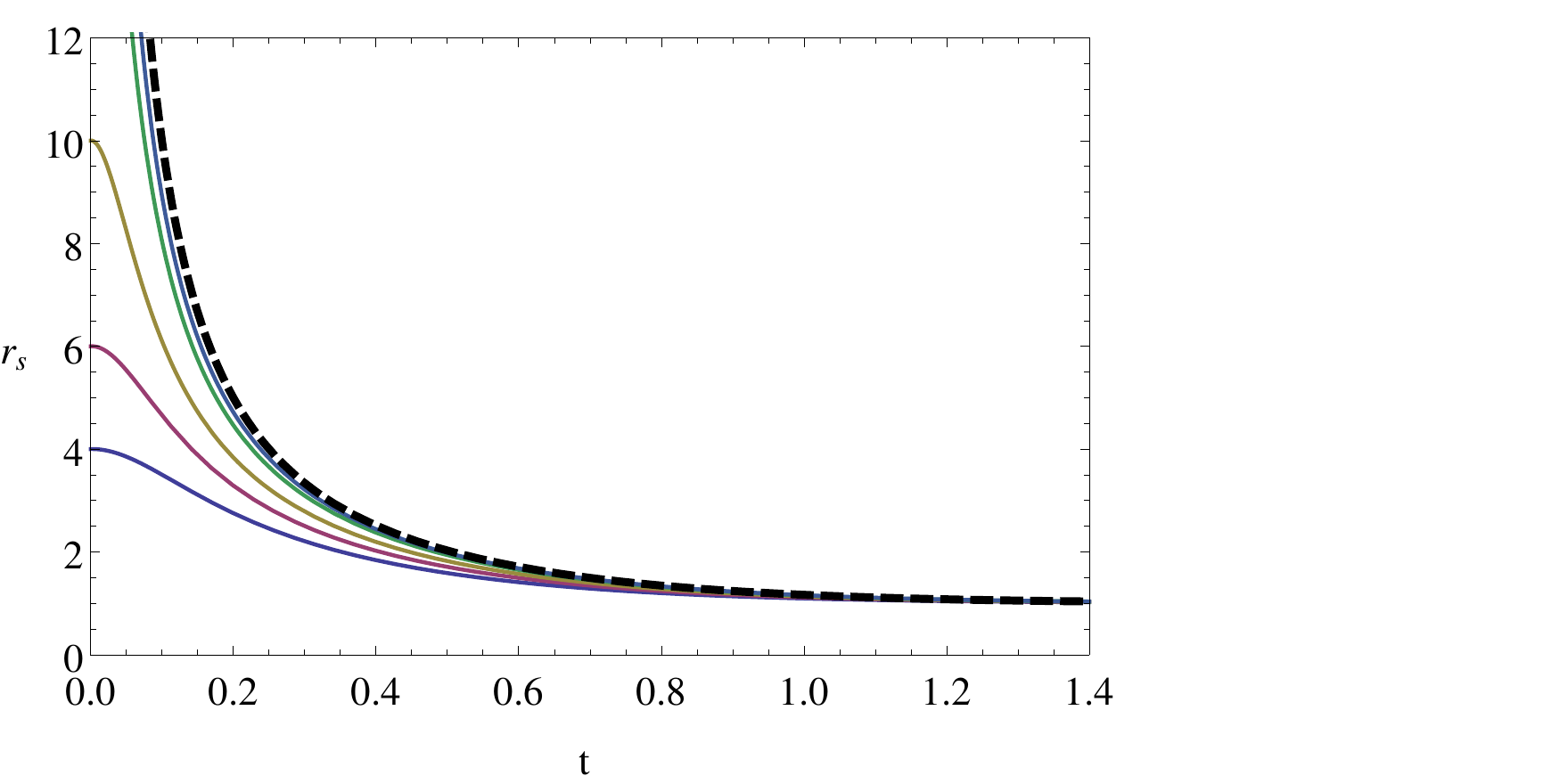}$\!\!\!\!\!\!\!\!\!\!\!\!\!\!\!\!\!\!\!\!\!\!\!\!\!\!\!\!\!\!\!\!\!\!\!\!\!\!\!$\includegraphics[scale=0.38]{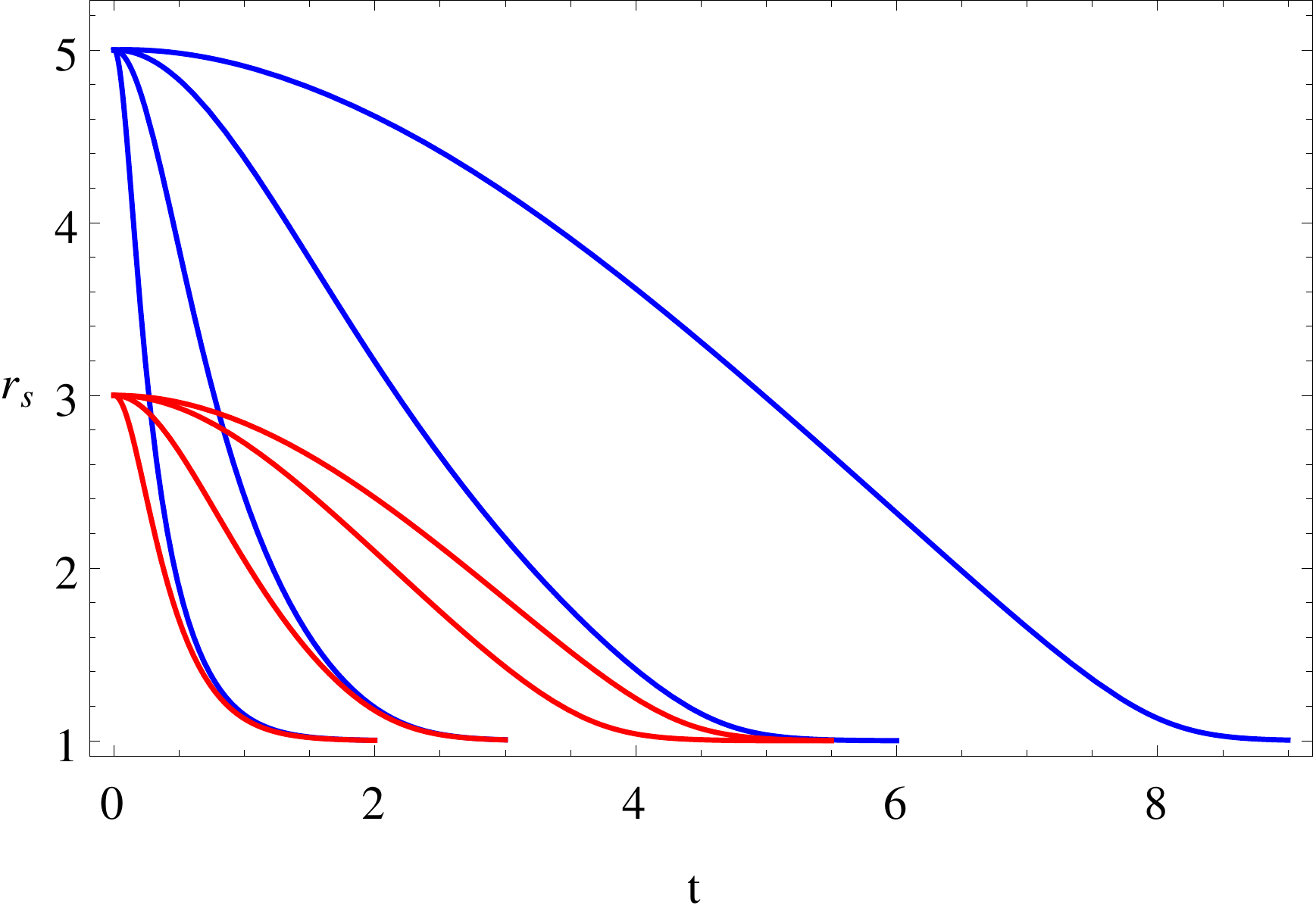}
\caption{\label{fig:shelltrajectories1} 
Left: Shell trajectories for $c=-1/3$ and for $r_0=(4,6,10,25,50)$ (bottom to top). The dashed curve represents here an ingoing null geodesic starting from the boundary at $t_+=0$.  Right: Shell trajectories for $r_0=3$ and 5 and for $c=0,$ 0.3, 0.33 and $1/3$ (from left to right). The units in both figures are chosen such that $r_h=1$.}
\end{center}
\end{figure}

Many of the interesting features of the shell trajectories become apparent only once the trajectory is expressed in terms of the time coordinate $t_{+}$. Solving eq.~(\ref{eq:tp}) leads to an expression for $t_{+}$ in terms of elliptic integrals, which is not particularly illuminating. We will thus rather take a step back and solve the EoM for $dr_s/dt_+$, obtained by taking the ratio of eqs.~(\ref{eq:rs}) and (\ref{eq:tp}),
\begin{equation}
\frac{dr_s}{dt_+}=\frac{\sqrt{r_0^4-r_s^4}\left(r_h^4-r_s^4\right)}{r_s^2\sqrt{r_0^4-r_h^4}}\, .
\end{equation}
Solving for $t_+$ from here, we obtain
\begin{equation}
t_+=\frac{\sqrt{r_0^4-r_h^4}}{r_0^3}\int_{r_s/r_0}^1\frac{du\,u^2}{\sqrt{1-u^4}\left[u^4-\left(\frac{r_h}{r_0}\right)^4\right]}\, ,\label{eq:tintegral1}
\end{equation}
where we have defined the integration variable $u=r/r_0$. A numerical integration of eq.~(\ref{eq:tintegral1}) is shown in figure \ref{fig:shelltrajectories1} (left). It is clearly seen from here that all trajectories asymptotically approach the horizon at $r=r_h=1$ with the same exponential rate as a null geodesic, but that the approach towards the null geodesic becomes faster when $r_0$ is increased. 

Both of the above features can be understood from the integral of eq.~(\ref{eq:tintegral1}).  Near the horizon, it is dominated by its lower limit, where we can approximate
\beq
t_+= \frac{1}{4r_h}\int_{r_s/r_0}\frac{du}{u-r_h/r_0}+...=-\frac{1}{4r_h}\log(r_s-r_h)+... \,,
\eeq
leading to the relation
\beq
r_s\approx r_h+C e^{-4 r_h  t_+}\,,
\eeq
i.e.~a null geodesic near the horizon. The exact same thing happens when $r_0$ is taken towards the boundary with $r_s/r_0$ fixed to a small number: The integral is again dominated by the lower limit of integration, and we can approximate
\beq
t_+=\frac{1}{r_0}\int_{r_s/r_0}\frac{du}{u^2}+...=\frac{1}{r_s}+... \, ,
\eeq
which gives
\beq
r_s\approx\frac{1}{t_+}\, ,
\eeq
identified as a null geodesic near the boundary. As $r_0\rightarrow\infty$, a boundary observer thus sees the shell apparoching a null geodesic, implying that the whole spacetime for $r\ll r_0$ is well approximated by the Vaidya limit. 

\subsection{Generic EoS: $-1<c<1/3$}
For the range $-1<c<1/3$, the shell trajectories share the same qualitative features as the above example $c=-1/3$; in particular, the they are always seen to approach a null geodesic when either $r_0/r_h\rightarrow\infty$ or $r_s\rightarrow r_h$. To demonstrate this, we work at the level of the EoM and show that it approaches the equation of a null geodesic,
\beq
\frac{dr}{dt_+}=-f_+(r)\, ,\label{eq:lightliketraj}
\eeq
in these limits. 

For a general $c$, the shell equations of motion are given by
\begin{align}
\dot{r}_s^2 &=- r_s^2 +\frac{r_h^4}{2\,r_s^2}+\frac{M^2}{4\,r_s^{4+6c}} + \frac{r_h^8 r_s^{6c}}{4\,M^2}\,,\label{eq:generalc1}
\\
\dot{t}_{\pm}&=\frac{\sqrt{f_{\pm}+\dot{r}_s^2}}{f_{\pm}} \,, \label{eq:generalc2}
\end{align}
where we will first consider the limit of the shell approaching the horizon, $r_s\rightarrow r_h$. In this case, $f_+$ approaches zero, so we can approximate (\ref{eq:generalc2}) as
\beq
\dot{t}_+\approx\frac{|\dot{r}_s|}{f_+}\,.
\eeq
Using this, we obtain
\beq
\frac{dr_s}{dt_+}=\frac{\dot{r_s}}{\dot{t}_+}\approx -f_+(r_s),
\eeq
which clearly implies that for all initial data with $r_0 > r_h$ the shell approaches the speed of light, as it approaches the horizon. 

Moving next to the case of $r_0/r_h\rightarrow\infty$, we substitute the integration constant
\beq
M\approx \frac{1}{2}r_h^4 r_0^{3c-1}
\eeq
into eq.~(\ref{eq:generalc1}). This leads to
\beq
\dot{r}_s^2 \approx - r_s^2 +\frac{r_h^4}{2\,r_s^2}+\frac{r_h^8 \, r_0^{6c-2}}{16 \, r_s^{6c+4}} + \frac{r_s^{6c}}{r_0^{6c-2}}\,,
\eeq
which shows that as long as $c<1/3$ and the shell location $r_s$ is kept fixed as $r_0/r_h\rightarrow\infty$, $\dot{r}_s\rightarrow\infty$ in this limit. Thus, in eq.~(\ref{eq:generalc2}) we can use $\dot{r}_s^2\gg f_+(r_s)$ and approximate
\beq
\dot{t}_{\pm}\approx \frac{|\dot{r}_s|}{f_{\pm}},
\eeq
which again leads to a null geodesic solution for $r_s$. 

One difference between the two limits we have been considering above is that as the starting point of the shell is sent to infinity, the shell trajectory approaches a null geodesic when viewed either from the outside of the shell, using the time coordinate $t_+$, or from the interior of the shell, using the time coordinate $t_-$. This is not the case when $r_0$ is kept fixed and the shell approaches the horizon.  In this case, only the trajectory as viewed from the exterior approches a null geodesic, while from the interior point of view it usually does not.

\subsection{Generic EoS: $c\ge 1/3$}

Finally, we take a look at values of $c$ greater or equal to the conformal value $c=1/3$. Again, the analysis of the trajectories near the horizon goes through unchanged, as the argument presented in the previous subsection is independent of the value of $c$. Thus, the shell is seen to approach the speed of light also for $c\ge 1/3$. 

The case of large $r_0$ is, however, very different. Here, the AdS spacetime can be seen to provide a harmonic potential that pulls the shell towards the center, seen in the first term of the right hand side of eq.~(\ref{eq:generalc1}). For $c>1/3$, the last term in this equation wins the pull of AdS, and the shell gets repelled from the center, accelerating towards the boundary. Physically, this means that the pressure of the shell wins over the gravitational attraction towards the center of AdS.

In the special case of $c=1/3$, the last term of eq.~(\ref{eq:generalc1}) also scales as $r_s^2$, which leads to the possibility that the ``forces'' cancel at large $r_s$. This enables the shell to approach the center of AdS with a very small acceleration even when it starts from near the boundary, as the gravitational pull and pressure almost cancel each other. Quantitatively, this can be seen by determining the acceleration of the shell,
\begin{equation}
\ddot{r}_s = \frac{1}{2\dot{r}_s}\frac{d}{d\tau} \dot{r}_s^2 \, ,
\end{equation}
and then expanding it near the turning point. Differentiating the expression in eq.~(\ref{shelleom1}) and expanding it yields
\begin{equation}
\ddot{r}_s = -4r_0 + 2\,(1+3c)\,r_0\,\sqrt{1-\frac{r_h^4}{r_0^4}} + \mathcal{O}(r-r_0) \; .
\end{equation}
From this expression, we see that the $c=\frac{1}{3}$ case is special, as for sufficiently large $r_0$ the acceleration vanishes to first order and the motion of the shell can be arbitrarily slow. In a spacetime of arbitrary dimenson $d+1$ (in our case $d=4$), the special value of $c$ reads $c=\frac{1}{d-1}$.

\section{Entanglement entropy \label{sec:entropy}}
Next, we move on to consider the covariant holographic entanglement entropy (HEE)  \cite{Hubeny:2007xt}, which is obtained by extremizing the area functional
\beq
A=\int d^3\sigma\sqrt{\det_{ab}g_{\mu\nu}\frac{\partial x^{\mu}}{\partial\sigma^a}\frac{\partial x^{\nu}}{\partial\sigma^b}}
\eeq
with the condition that the bulk hypersurface ends on a predefined surface $\mathcal{A}$, which resides on a constant time slice on the boundary. In the dual CFT, the (geometric) entanglement entropy of the region $\mathcal{V}$, whose boundary $\mathcal{A}$ is, is then conjectured to be given by
\beq
S_\text{EE}=\frac{A}{4G_N}\, .
\eeq
The time evolution of this quantity has been extensively analyzed in various equilibration scenarios since the original work of \cite{Hubeny:2007xt}; in particular, for studies in the Vaidya spacetime, see \cite{AbajoArrastia:2010yt,Albash:2010mv,Balasubramanian:2011ur}.

In the current section, we will consider the evolution of the HEE in the collapsing shell model, using the physical shell trajectories obtained in the previous section. First, we study a simple example shape for the boundary region, a strip of width $L$, and then derive some more generic results for arbitrary shapes. To supplement this discussion, the relevant equations of motion for the case of a spherical boundary region are derived in some detail in appendix \ref{sec:spherical}. 

In all of the calculations presented in this section, we work in the Eddington-Finkelstein coordinates, where instead of using the time coordinate $t$ we employ the lightcone coordinate $v_{\pm}$, defined by
\beq
dv_{\pm}=dt_{\pm}+\frac{dr}{f_{\pm}(r)}\,.
\eeq
In addition, we will switch to the bulk radial coordinate $z=1/r$, so that our bulk metric will be given by
\beq
ds^2=\frac{1}{z^2}\left[-h(z,v)dv^2-2dv dz+d\textbf{x}^2\right]\, .
\eeq
Here, we have further defined
\beq
h(z,v)=1-\theta(v-v_s(z))z^4\,,
\eeq
where $v_s(z)$ is the trajectory of the shell parametrized as a function of $z$. In this entire section, we set the Schwarzschild radius to unity, i.e.~$r_h=1/z_h=1$.

\subsection{Strip boundary region}
The interior of a strip on the boundary is defined as the region of space with $x^1\in (-L/2,L/2)$, $x^2\in (-L_2/2,L_2/2)$, and $x^3\in (-L_3/2,L_3/2)$, where $L_2$ and $L_3$ will be sent to infinity at the end. In this case, we can clearly assume that the bulk extremal surface is invariant under translations in the $x^2$ and $x^3$ directions. Thus, we can parametrize the extremal surface using the coordinates $z=z(x)$ and $v=v(x)$, where $x\equiv x^1$, while the surface is spread homogenously along the $x^2$ and $x^3$ directions. For a more thorough explanation of the setup, we refer the reader to \cite{Albash:2010mv,Balasubramanian:2011ur}. 

With the above definitions, the area functional under consideration becomes
\beq
A=L_2L_3\int dx\frac{\sqrt{B}}{z(x)^3}=L_2L_3\int dx\mathcal{L},\quad B=1-h(z(x),v(x))\,v'(x)^2-2\,v'(x)z'(x)\,.
\eeq
Due to the translational invariance of the system, there is a conserved quantity
\beq
\frac{\partial \mathcal{L}}{\partial z'}z'+\frac{\partial \mathcal{L}}{\partial v'}v'-\mathcal{L}=-\frac{1}{z^3\sqrt{B}}\, ,
\eeq
which is indeed constant along the entire extremal surface. Its value can be fixed by evaluating it at the point $(z_*,v_*,x_*)$ where the surface turns around,\footnote{Here we 
assume that the extremal surface is reflection symmetric around 
$x=0$.}, i.e.~$z'(x_*)=v'(x_*)=0$ which quickly leads to the result
\beq
\sqrt{B}=\left(\frac{z_*}{z}\right)^3\,.\label{eq:xconstant}
\eeq

The metric is clearly independent of $v$ everywhere except at the position of the shell, implying that there is also a second constant of motion,
\beq
\frac{\partial\mathcal{L}}{\partial v'}=-\frac{h v'+z'}{z^3\sqrt{B}}\equiv -\tilde{E}\,,
\eeq
which takes different values on the two sides of the shell. Using eq.~(\ref{eq:xconstant}), we obtain
\beq
h_{\pm}v'+z'=E_{\pm}\,,\label{eq:energy}
\eeq
where we have redefined the constant as $E_{\pm}\equiv z_*^3\tilde{E}_{\pm}$, and denoted $h_-=1$ and $h_+=1-z^4$. Solving this equation for $v'$ and plugging the result into eq.~(\ref{eq:xconstant}) finally leads us to
\beq
z'^2=E_{\pm}^2+h_{\pm}\left[\left(\frac{z_*}{z}\right)^6-1\right]\equiv H_{\pm}(z)\,.\label{eq:zdiffeq}
\eeq

If the boundary separation $L$ is sufficiently small, the extremal surface never reaches the shell and always stays in the black hole region, implying that the entanglement entropy stays thermal at all times. The precise value of $L$, above which the surface crosses the shell, clearly depends both on the trajectory of the shell and on the boundary time. In the rest of this section, we will only consider the interesting case, where $L$ is large enough so that the surface crosses the shell in the beginning of the time evolution.

We start by studying the equations of motion in the pure AdS region inside the shell, where the extremal surfaces always have a turning point with $z'=v'=0$. It is easy to see that $E_-$ vanishes there, which implies that everywhere inside the shell we have
\beq
v'=-z'\,.\label{eq:constanttime}
\eeq
Written in terms of the $t_-$ coordinate, this means that $t'_-=0$, i.e.~that the surface lies in a constant time slice inside the shell. Integrating eq.~(\ref{eq:zdiffeq}) is now a straighforward excercise, and one finds a two parameter family of solutions parameterized by the turning point $z_*$ and the value of the time coordinate there, $v(z_*)$. The latter of these parameters can, however, be further traded for the point $z_c$ where the extremal surface crosses the shell, so that the interior surface is parameterized by the pair $(z_*,z_c)$. In the following, we will need the value of the derivative at the interior  shell position $z'_-\equiv z'(x_c)$, which is given by (cf.~eq.~(\ref{eq:zdiffeq}))
\beq\label{inside}
z_-'=-\frac{1}{z_c^3}\sqrt{z_*^6-z_c^6} \,.
\eeq

Next, we continue the extremal surfaces across the shell using the junction conditions of eqs.~(\ref{eq:match1}) and (\ref{eq:match2}), with $\sigma=x$. In our current coordinate system, these read
\begin{align}
z'_+&=\frac{1}{z_c^2}\left[\alpha_+\alpha_-+\dot{z}_s(\alpha_--\alpha_+)-\dot{z}_s^2\right]z'_-
+\frac{\dot{z}_s}{z_c^2}(\alpha_--\alpha_+)v'_- \,,
\\
v'_+&=\frac{1}{z_c^2h_+(z_c) }\left[\alpha_+\alpha_--\dot{z}_s(\alpha_--\alpha_+)-\dot{z}_s^2\right]v'_-\,,
\end{align}
where $z_{\pm}$ and $v_{\pm}$ are the corresponding derivatives evaluated on the outside and inside of the shell, $\alpha_{\pm}\equiv\sqrt{h_{\pm}z_c^2+\dot{z}_s^2}$, and $\dot{z}_s$ is the proper velocity of the shell. Inside the shell, we can further use eq.~(\ref{eq:constanttime}) to write $v'_-=-z'_-$, which reduces the junction conditions to
\begin{align}
\frac{z'_+}{z'_-}&\equiv Z(\dot{z}_s)=\frac{1}{z_c^2}\left(\alpha_+\alpha_--\dot{z}_s^2\right) \,,\nonumber
\\
\frac{v'_+}{z'_-}&\equiv V(\dot{z}_s)=-\frac{1}{z_c^2h_+(z_c)}\left[\alpha_+\alpha_--\dot{z}_s(\alpha_--\alpha_+)-\dot{z}_s^2\right]\,.\label{eq:matching}
\end{align}
As a sidenote, we remark that we have here introduced a notation, where the junction conditions are considered functions of the derivative terms; this is done in anticipation of the following section, where we will consider the effects of the quasistatic approximation where these derivatives are altogether ignored.

From eq.~(\ref{eq:matching}), we see that the quantity $z'_-$, which depends on $z_*$ and $z_c$, determines the values of the derivatives $z'_+$ and $v'_+$ outside the shell. Thus the integration constant $E_+$ gets determined by $z_*$ and $z_c$ using eq.~(\ref{eq:energy}), and we can therefore denote $E_+=E_+(z_*,z_c)$. Now the boundary quantities can also be straightforwardly calculated using eqs.~(\ref{eq:energy}) and (\ref{eq:zdiffeq}), and in particular the length of the boundary interval becomes
\beq
L/2=\int_{z_c}^{z_*}\frac{dz}{\sqrt{H_-(z)}}+\int_{z_c}^{z_{max}}\frac{dz}{\sqrt{H_+(z)}}+\int_{0}^{z_{max}}\frac{dz}{\sqrt{H_+(z)}}\,.\label{eq:L}
\eeq
Here, we have denoted by $z_{max}$ the maximal value that the coordinate $z$ obtains along our extremal surface within the outside region. If $z'_+<0$, then $z_{max}=z_c$ as the surface climbs monotonically up towards the boundary, while if $z'_+>0$, then $z_{max}$ is the point at which the surface turns around outside the shell, to be determined from the condition
\beq
H_+(z_{max})=0.
\eeq
The time, at which the surface reaches the boundary, is on the other hand given by
\beq
t=v_s(z_c)+\int_{z_c}^{z_{max}}\frac{dz}{h_+}\Bigg[\frac{E}{\sqrt{H_+(z)}}-1\Bigg]+\int_{0}^{z_{max}}\frac{dz}{h_+}\Bigg[\frac{E}{\sqrt{H_+(z)}}+1\Bigg]\,,\label{eq:t}
\eeq
while the area of the extremal surface reads
\beq
A=2L_2L_3z_*^3\Bigg[\int_{z_c}^{z_*}\frac{dz}{z^6\sqrt{H_-(z)}}+\int_{z_c}^{z_{max}}\frac{dz}{z^6\sqrt{H_+(z)}}+\int_0^{z_{max}}\frac{dz}{z^6\sqrt{H_+(z)}}\Bigg]\,.\label{eq:a}
\eeq
As we can see from here, all boundary quantities have now been given implicitly in terms of two parameters: the turning point $z_*$ and the crossing location $z_c$.

\subsubsection{Early time behavior}

At early times, right after the shell is released from rest, the geometry is close to being static, and we can work in an expansion around a static shell. The relevant extremal surfaces then lie close to constant $t$ surfaces, making it appropriate to use the $(z,t)$ coordinate system. In particular, the shell trajectory near the turning point can be written in terms of a proper acceleration $a=\ddot{z}_s(0)$ as
\beq
z_s(\tau)=z_0+\frac{1}{2}a\tau^2+O(\tau^3)\,,
\eeq
where $a$ can be determined from the equation of motion of the shell. 

Expanding now the junction conditions to first order in powers of $\dot{z}_s=a\tau+O(\tau^2)$ gives
\beq
z'_+=\sqrt{h_+}z'_-,\quad t'_+=\frac{1-\sqrt{h_+}}{z_ch_+}\dot{z}_s z'_-\,.
\eeq
The boundary length and the area of the extremal surface are again given by eqs.~(\ref{eq:L}) and (\ref{eq:a}), while the boundary time becomes
\beq
t=t_s(z_c)+\int_0^{z_c}dz\frac{E_+}{h_+(z)\sqrt{H_+(z)}}\,.\label{eq:t2}
\eeq
Here, $t_s(z_c)$ is once again the shell trajectory, now parametrized in terms of $z$ and evaluated at the point
where the extremal surface crosses the shell, $z=z_c$. In the following we will for simplicity denote $t_s(z_c)\equiv t_c$. Finally, the proper time $\tau$ can at early times be approximated by the proper time measured by an observer at rest at $z=z_c$,
\beq
\tau\approx \sqrt{h_+(z_c)}t_c/z_c\,.
\eeq

At this point, an important observation is that $t'_+$ is proportional to $\dot{z}_s\approx a\tau$, which is small at early times. Being proportional to $t'_+$, $E_+$ is therefore also small, and we can expand eqs.~(\ref{eq:L}), (\ref{eq:a}) and (\ref{eq:t2}) in powers of $E_+$ and $\delta z\equiv z_c-z_0=a\tau^2/2$. Assuming $L$ to be large, so that $z_*$ has to be sizable as well, we obtain for $L$
\beq
\frac{L}{2}=z_* \sqrt{\pi}\frac{\Gamma\left(\frac{2}{3}\right)}{\Gamma\left(\frac{1}{6}\right)}+O(z_*^{-3})+O(z_*^{-3}\tau^2)\,.
\label{eq:largezstar}
\eeq
We see from here that $z_*$ varies in time only very slowly, as the time dependence is supressed by an overall factor $z_*^{-3}$. Thus, $z_*$ is fixed in terms of $L$. 

\begin{figure}[t]
\begin{center}
\includegraphics[scale=.7]{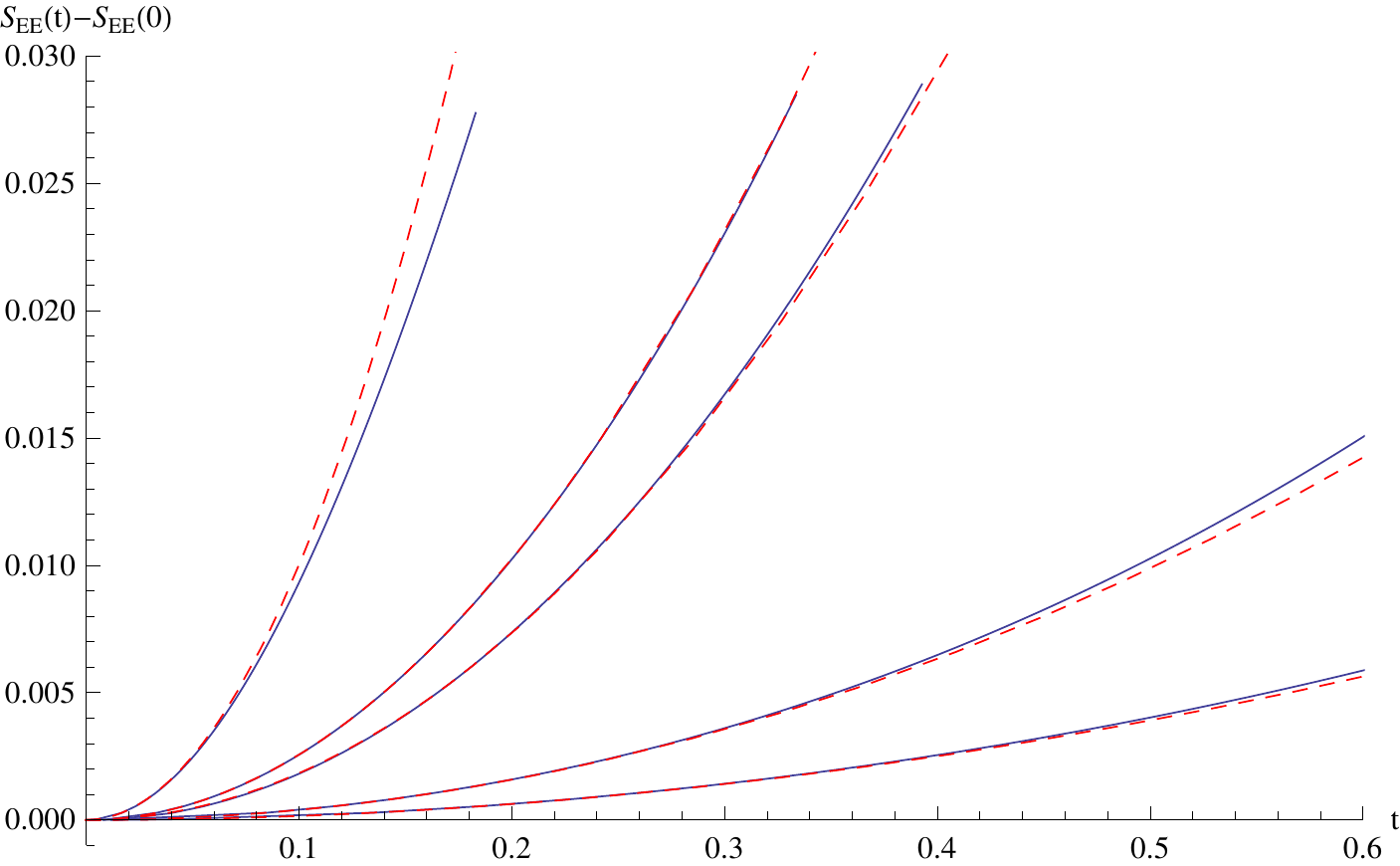}
\caption{\label{fig:earlytimeanalytic} The time dependence of the HEE (solid blue curves), obtained through a numerical integration of the extremal surface equations of motion, compared to the early time analytic solution of eq.~(\ref{analyticres}) (dashed red curves). Here $z_0=0.5$ and $L=8$, while $c=(-1,0,0.1,0.3,1/3)$ (left to right). }
\end{center}
\end{figure}

To first order in $E_+$, the boundary time becomes now
\beq
t=t_c+\int_0^{z_0}dz\frac{E_+}{h_+\sqrt{H_0(z)}}+O(\tau^3)\,,
\eeq
where $H_0(z)=H_+(z)-E_+^2$. For large $z_*$, the integral can be easily evaluated, leading to
\beq
t\approx t_c\left[1-\frac{(1-\sqrt{h_+(z_0)})^2}{2 z_0^5}a\right]\,,\label{eq:boundaryt}
\eeq
where we have used
\beq
E_+=-\frac{\sqrt{h_+(z_c)}(1-\sqrt{h_+(z_c)})}{z_c^5}z_*^3 a t_c 
\eeq
and replaced $z_c$ by $z_0$, which is allowed to leading order in $\tau$. Similarly, we can expand the area in eq.~(\ref{eq:a}) in powers of $E_+$ and $\delta z$, which leads to the change in the area equaling
\beq
\Delta A=A(t)-A(t=0)= L_2 L_3\frac{\sqrt{h_+(z_0)}(1-\sqrt{h_+(z_0)})}{z_0^{5}} at_c^2\left[1-\frac{(1-\sqrt{h_+(z_0)})^2}{2z_0^{5}}a\right]\,.\nonumber
\eeq

Combining finally the above results produces
\beq
\Delta A= \frac{1}{2}A_{\partial A}\frac{\sqrt{h_+(z_0)}\left(1-\sqrt{h_+(z_0)}\right)}{z_0^{5}}\frac{a t^2}{\left[1-\frac{(1-\sqrt{h_+(z_0)})^2}{2z_0^{5}}a\right]}\,, \label{analyticres}
\eeq
where we denote the area of the boundary entangling surface as $A_{\partial A}=2L_2L_3$. This formula nicely demonstrates the relation between the entanglement growth and acceleration at early times. The early time behavior obtained here is compared to a full numerical integration of eqs.~(\ref{eq:L}), (\ref{eq:t}) and (\ref{eq:a}) in fig.~\ref{fig:earlytimeanalytic}, which shows an impressive agreement up to relatively large time scales.

\subsubsection{Linear scaling}

For large values of $L$, the quadratic early time behaviour of the entanglement entropy is followed by a long regime of linear increase, where $\Delta S_\text{EE}\sim \Delta t$. In the case of Vaidya collapse, the existence of this region was demonstrated in \cite{Liu:2013qca}, where it was seen to emerge from extremal surfaces inside the horizon of the black hole. In what follows, our analysis is closely related to the Vaidya case, and to this end we refer the interested reader to \cite{Liu:2013qca} for more details. Our goal will be to  provide a simple and hopefully intuitive picture of where and why the linear region appears in our setup, highlighting the main differences that arise due to the slower motion of the shell. The precise details of the shell motion are unimportant for what follows, so we will only use the fact that $z_s(v)$ is a monotonically increasing function of $v$ and that the shell does not move faster than the speed of light. A key assumption in our calculation is that the relevant extremal surfaces at late times are those that pass through the black hole horizon, which can indeed be shown to be true by numerically constructing the relevant surfaces. 

\begin{figure}[t]
\begin{center}
\includegraphics[scale=1.2]{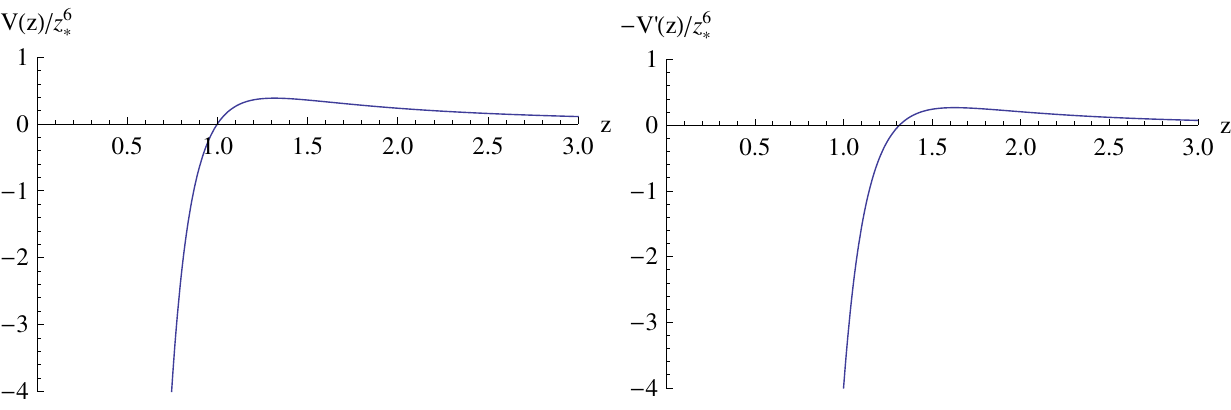}
\caption{\label{fig:potential} The potential in the effective particle problem. Note that in this figure the boundary resides at $z=0$, while the black hole singularity lives at $z=\infty$.}
\end{center}
\end{figure}

The equation of motion of extremal surfaces in the black hole region, cf.~eq.~(\ref{eq:zdiffeq}), can be written in a suggestive form
\beq
z'^2+V(z)=\mathcal{E},\quad V(z)=z_*^6\left(\frac{1}{z^2}-\frac{1}{z^6}\right),\quad \mathcal{E}=E_+^2,
\eeq
where we have neglected two terms subleading at large $z_*$. In the following, we will think of this equation as an EoM for a non-relativistic particle moving in the potential $V(z)$, with $x$ interpreted as a fictitious time coordinate. This potential is plotted in fig.~\ref{fig:potential}.

The rule for constructing the extremal surface is  as follows. As we saw previously, just outside the shell the derivatives $z'$ and $v'$ are determined by the two parameters $z_*$ and $z_c$ through the junction conditions of eq.~(\ref{eq:matching}). This setup is clearly equivalent to a classical mechanics problem with a particle starting from $z=z_c$ with some fixed initial velocity $z'=v_0$ and with the requirement of having to end up at the boundary. As can be seen from figure \ref{fig:potential}, the potential has a maximum at some $z=z_m$, and for $z>z_m$ the force felt by the particle tries to pull it towards the singularity. The location of this point is given by
\beq
V'(z_m)=0 \quad \Rightarrow \quad z_m=3^{1/4}.
\eeq

There are clearly two ways to avoid the fall of the particle into the black hole singularity. The first one is to have it start from $z_c > z_m$ and give it enough negative initial velocity to get over the potential barrier at $z_m$. This is, however, not allowed by the junction conditions, which determine the initial velocity through 
\beq
v_0=-\frac{z_*^3}{z_c^5}A(z_c),\quad A(z_c)=\sqrt{z_c^2h(z_c)+\dot{z}_s^2}\sqrt{z_c^2+\dot{z}_s^2}-\dot{z}_s^2\,.\label{eq:v0}
\eeq
For general real values of $\dot{z}_s$, the function $A(z_c)$ takes negative values in the region $z_c>\bar{z}_c$, where $1<\bar{z_c}<2^{1/4}$ and where the upper bound is approached when we approach the Vaidya spacetime as $\dot{z}_s\rightarrow \infty$. This means that $A(z_c)$ is negative and $v_0$ positive for $z_c\ge z_m$, which implies that the particle will unavoidably fall into the singularity.

A second way to reach the boundary is to start from $z<z_m$ and choose the initial velocity to be either positive but sufficiently small so that the particle will not get over the potential barrier, or alternatively even negative. For reasons that will become clear in a moment it is the first case that turns out to be the relevant one for us. Then, the maximum value for the energy of the particle is
\beq
\mathcal{E}_{max}=V(z_m)=-\frac{z_*^6h(z_m)}{z_m^6}\,,
\eeq
so that $E_+$ should be bounded by $\sqrt{\mathcal{E}_{max}}$. In order to make the initial velocity $v_0$ given by eq.~(\ref{eq:v0}) small, we must obviously have the coefficient $A(z_c)$ be very small. As $A(z_c)$ changes sign at $\bar{z}_c$, which falls in the interval $(1,z_m)$, we can do this by choosing $z_c$ to be sufficiently close to $\bar{z}_c$. 

Since we want the value of $v$, at which the particle reaches the boundary, to be large, it should spend a long fictitious time $x$ in the bulk. After the above considerations, it should now be clear how to arrange this: We should choose the initial velocity to be such that the particle almost reaches $z_m$ and then turns around. To quantify this statement, we expand the potential around $z=z_m$, obtaining
\beq
z'^2-\omega^2(z-z_m)^2 =-\omega^2\delta\mathcal{E},\quad \omega^2=-\frac{1}{2}V''(z_m),\quad \delta \mathcal{E}=(\mathcal{E}_{max}-\mathcal{E})/\omega^2\,, \label{eq:expandedeom}
\eeq
and then integrate this expression to get
\beq
\Delta x\approx \frac{2}{\omega}\int_{\sqrt{\delta\mathcal{E}}}\frac{dy}{\sqrt{y^2-\delta\mathcal{E}}}=\frac{1}{\omega}\log\left(\frac{\mathcal{E}_{max}-\mathcal{E}}{\omega^2}\right)\,.\label{eq:harmonicintegral}
\eeq
Here, we have used the integration variable $y\equiv z_m-z$ and neglected the contribution of the upper integration limit, as it is subleading in the limit of interest, $\delta\mathcal{E}\rightarrow 0$. The factor of 2 in front of the integral is due to the symmetricity of the trajectory. 

The value of the boundary time coordinate of the extremal surface is determined by
\beq
v'=\frac{\sqrt{\mathcal{E}}-z'}{h}\,.
\eeq
As $z'$ is very small for most of the time, we can treat the right hand side of this equation as a constant, producing
\beq
t\approx -\frac{\sqrt{\mathcal{E}_{max}}}{h(z_m)}\Delta x.
\eeq
The area of the extremal surface is then given by
\beq
A\approx \frac{4 L_2L_3z_*^3}{z_m^6\omega}\int_{\sqrt{\delta\mathcal{E}}}\frac{dy}{\sqrt{y^2-\delta\mathcal{E}}}= \frac{2 L_2L_3z_*^3}{z_m^6}\Delta x\,,
\eeq
from which we obtain, using $\Delta x=-t\frac{h(z_m)}{\sqrt{\mathcal{E}_{max}}}$ and $\mathcal{E}_{max}=-\frac{z_*^6h(z_m)}{z_m^6}$,
\beq
A\approx 2 L_2L_3  \,\frac{\sqrt{-h(z_m)}}{z_m^3}\,t\,.
\eeq

Inspecting the obtained result, we clearly observe that the area increases linearly in time. Following \cite{Liu:2013qca} in defining $v_\text{E}=\sqrt{-h(z_m)}/z_m^3$ as well as in denoting the thermal entropy density by $s_{th}=1/(4 G_N)$ and the area of the boundary entangling region by $A_{\partial A}=2 L_2 L_3$, we can now write
\beq
S_\text{EE}\approx A_{\partial A}s_{th}v_\text{E} t\,,
\eeq
where the numerical value of $v_\text{E}$ is found to be
\beq
 v_\text{E}=\frac{\sqrt{2}}{3^{3/4}}\approx 0.620403\,.
 \eeq
This is exactly the result obtained for the Vaidya case in \cite{Liu:2013qca}. The above discussion should make it clear that the result is independent of the shell trajectory and on the precise form of the junction conditions. 

\subsection{Linear scaling for general shapes}

In ref.~\cite{Liu:2013qca}, the authors argue that in order to isolate the linear regime in the time evolution of the HEE, one can take a limit in which turning point of the geodesic is sent to infinity before considering the time to be large. In this limit, the time evolution of the extremal surface is mostly due to the region inside the black hole horizon but outside the shell (i.e.~within the AdS Schwarzschild metric), and the surface furthermore moves very little in the $x$ direction in comparison with the part in the pure AdS region. To obtain the leading contribution to the area of the surface, it is therefore sufficient to approximate it as moving only in the $v$ direction. This way, the area functional becomes
\beq
A=A_{\partial A}\int dz\frac{1}{z^3}\sqrt{-h \Big(\frac{dv}{dz}\Big)^2-2\frac{dv}{dz}}\,,
\eeq
where $A_{\partial A}$ is again the area of the boundary theory entangling region and we have parameterized the surface with $v=v(z)$, denoting $v'\equiv dv/dz$. 

Within the black hole region, there is again a conserved ``Hamiltonian''
\beq
\frac{\partial L}{\partial v'}=-\frac{hv'+1}{z^3\sqrt{-h (v')^2-2v'}}=-E\,,
\eeq
from which we can easily solve
\beq
v'=\frac{1}{h}\left(\pm\frac{E}{\sqrt{E^2+\frac{h}{z^{6}}}}-1\right)\, ,
\eeq
where the $\pm$ correspond to two branches of solutions. Upon integration, this finally gives
\beq
t=v_s(z_c)-\int_{z_c}^{z_{max}}\frac{dz}{h}\left(\frac{E}{\sqrt{E^2+\frac{h}{z^{6}}}}+1\right)-\int_{0}^{z_{max}}\frac{dz}{h}\left(\frac{E}{\sqrt{E^2+\frac{h}{z^{6}}}}-1\right)\,.
\eeq
Here, the trajectory of the shell appears only via $v_s(z_c)$, which is obtained by inverting the function $z_s(v)$. 

Next, let us concentrate on the dominant contribution to the above $z$ integral that originates from the region near $z=z_{max}$. As the first two terms of the Taylor expansion of $\sqrt{E^2+\frac{h}{z^{6}}}$ aroud this point vanish, the value of $t$ diverges logarithmically, signaling that these specific values of $E$ and $z_{max}$ correspond to a critical surface. This way, we are lead to the two conditions
\beq
E^2+\frac{h(z_{max})}{z_{max}^{6}}=0,\quad \frac{\partial}{\partial z_{\max}}\frac{h(z_{max})}{z_{max}^{6}}=0\,,
\eeq
of which the second one can be solved for the value of $z_{max}$ and the first one for $E$, giving
\beq
z_{max}=z_m=3^{1/4},\quad E=\sqrt{-\frac{h(z_m)}{z_m^{6}}}\,.
\eeq
Using these results, the area becomes
\beq
A=A_{\partial A}\int_{z_c}^{z_{max}}\frac{dz}{z^{6}}\frac{1}{\sqrt{E^2+\frac{h}{z^{6}}}}+A_{\partial A}\int_{0}^{z_{max}}\frac{dz}{z^{6}}\frac{1}{\sqrt{E^2+\frac{h}{z^{6}}}}\,,
\eeq
and matching the main logarithmic contributions of the two integrals produces
\beq
A\approx A_{\partial A}\frac{-h(z_m)}{E z_m^{6}}t\,.
\eeq
Finally, we approximate $E$ by its value at the critical surface and thereby obtain
\beq
A\approx A_{\partial A} v_\text{E} t\, , \quad
v_\text{E}=\sqrt{-\frac{h(z_m)}{z_m^{6}}} \, ,
\eeq
a result in full agreement with that of \cite{Liu:2013qca}.

\section{Comparison with common approximation schemes}

The collapsing shell model of gauge theory thermalization has been extensively used not only in the context of studying entropies, but also to evaluate various correlation functions, corresponding e.g.~to the electromagnetic current operator or the energy momentum tensor on the field theory side \cite{Lin:2008rw,Baier:2012tc,Baier:2012ax,Steineder:2012si,Steineder:2013ana,Stricker:2013lma}. While entanglement entropy calculations such as \cite{AbajoArrastia:2010yt,Albash:2010mv,Liu:2013iza,Liu:2013qca,Balasubramanian:2011ur} are typically performed in the Vaidya limit of a lightlike shell, the Green's functions are usually determined in the opposite `quasistatic' approximation, in which the shell is taken to be a static object when formulating the junction conditions for the corresponding bulk fields. Physically, this approximation amounts to assuming that the time scale associated with the collapse of the shell is considerably larger than any other time scales relevant for the system, such as the inverse energy scale $1/\omega$ of the two-point function considered.

The calculations we have performed in this paper for the HEE allow us to make an interesting comparison between our `exact' results, derived for shells following realistic trajectories and employing the full junction conditions, and the corresponding quasistatic and Vaidya limits thereof. To obtain the former limit, we simply set $\dot{z}_s=0$ in the junction conditions of eq.~(\ref{eq:matching}), reducing them to the simple forms
\begin{eqnarray}
\frac{z'_+}{z'_-}&=& Z(0)=\frac{\alpha_+\alpha_-}{z_c^2} \,,\nonumber \\
\frac{v'_+}{z'_-}&=& V(0)=-\frac{\alpha_+\alpha_-}{z_c^2h_+(z_c)}\,,\label{qs}
\end{eqnarray}
where $z_-'$ is given by eq.~(\ref{inside}).\footnote{The physical timelike trajectories are, however, used elsewhere in the calculation.} At the same time, the Vaidya result is available by merely replacing the shell trajectory by an ingoing lighlike geodesic. Naively, we expect the quasistatic approximation to be valid only at the earliest times, i.e.~near the turning point of the shell, while the Vaidya limit should be approached at late times.

\begin{figure}[t]
\begin{center}
\includegraphics[scale=.4]{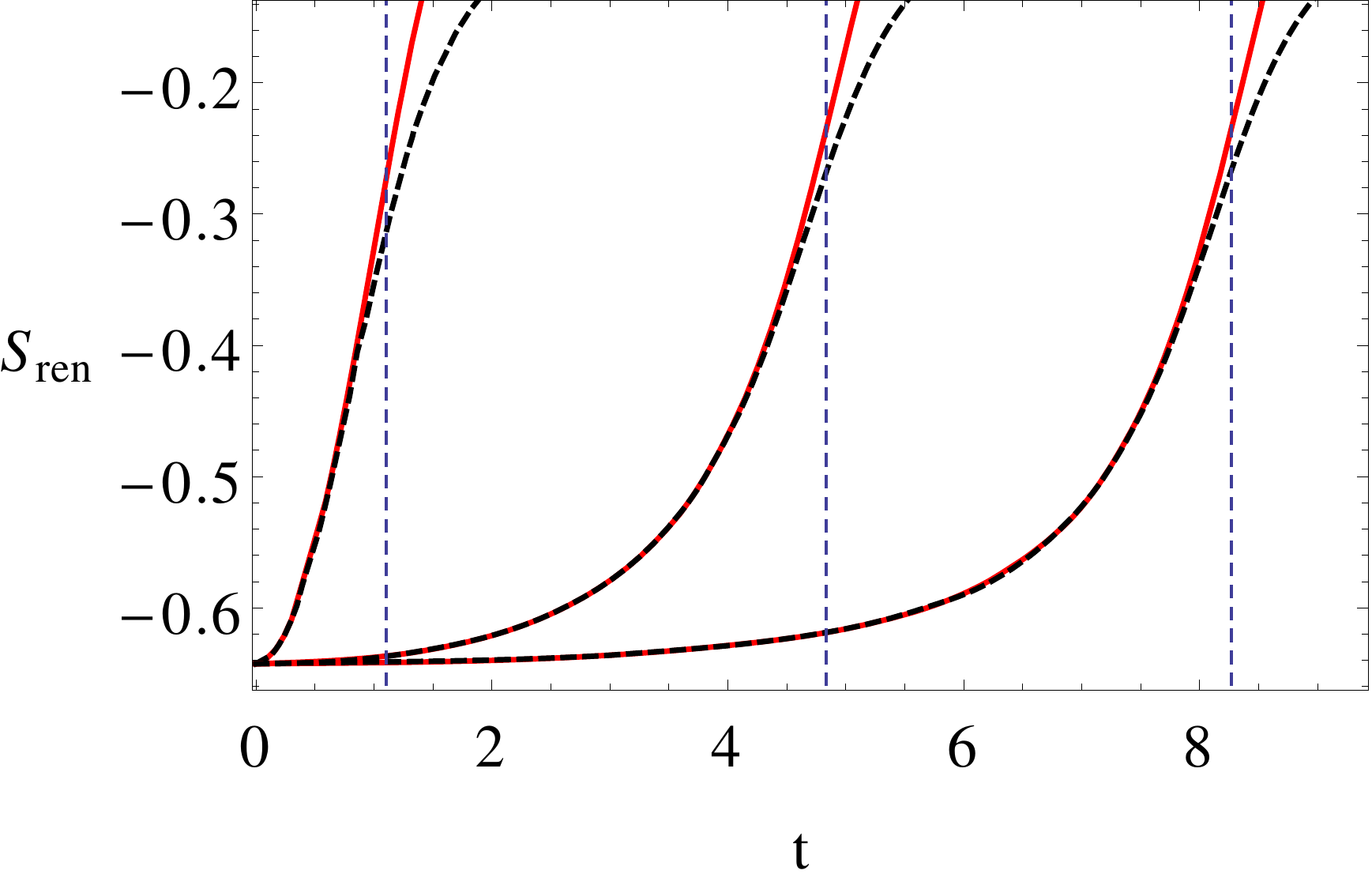}$\;\;\;\;\;$\includegraphics[scale=.4]{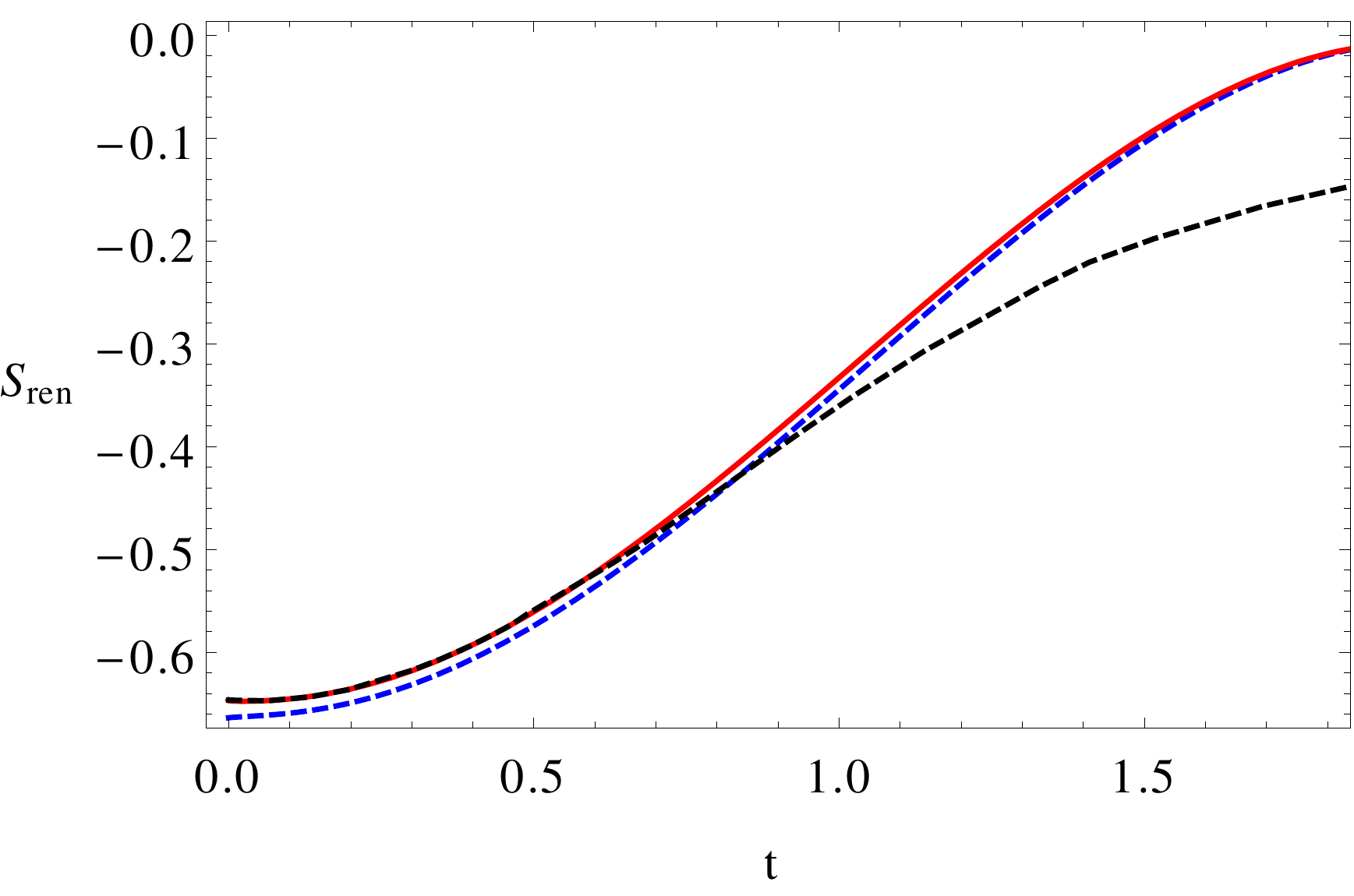}
\caption{\label{fig:staticapprox} 
Left: The HEE evaluated for a sphere of radius $R=2$ for $z_0=0.2$ and $c=0,\;0.33,\;1/3$ (from left to right). Shown here are the full results (solid red lines) as well as the quasistatic (black dashed lines) approximation thereof. Right: A zoom-in to the $c=0$ case, with the Vaidya limit (blue dotted line) added in the plot. 
 }
\end{center}
\end{figure}

In figure \ref{fig:staticapprox}, we display the result of the above comparison for the HEE evaluated for a sphere of radius $R=2$. Somewhat to our surprise, we observe from the left figure that independent of the value of $c$, the static approximation appears to work rather well until fairly late times and only breaks down when the shell is very close to the horizon; to aid the comparison, we have marked with dashed vertical lines the boundary times, at which those extremal surfaces that intersect the shell very close to the horizon, at $z_s=0.99z_h$, anchor to the boundary. At this point, the deviation of the quasistatic result from the full one is still only at the $10\%$ level. In the right figure, we take a closer look at the $c=0$ case and include in the figure also the Vaidya limit. We observe that the Vaidya result in turn gives a very good approximation of the full one already at relatively early times, and in particular that a combination of the quasistatic and Vaidya curves approximates the physical behavior of the entropy to a few percent level at all times. It is tempting to speculate, whether this observation would generalize to other, more complicated observables as well.

\begin{figure}[t]
\begin{center}
\includegraphics[scale=.55]{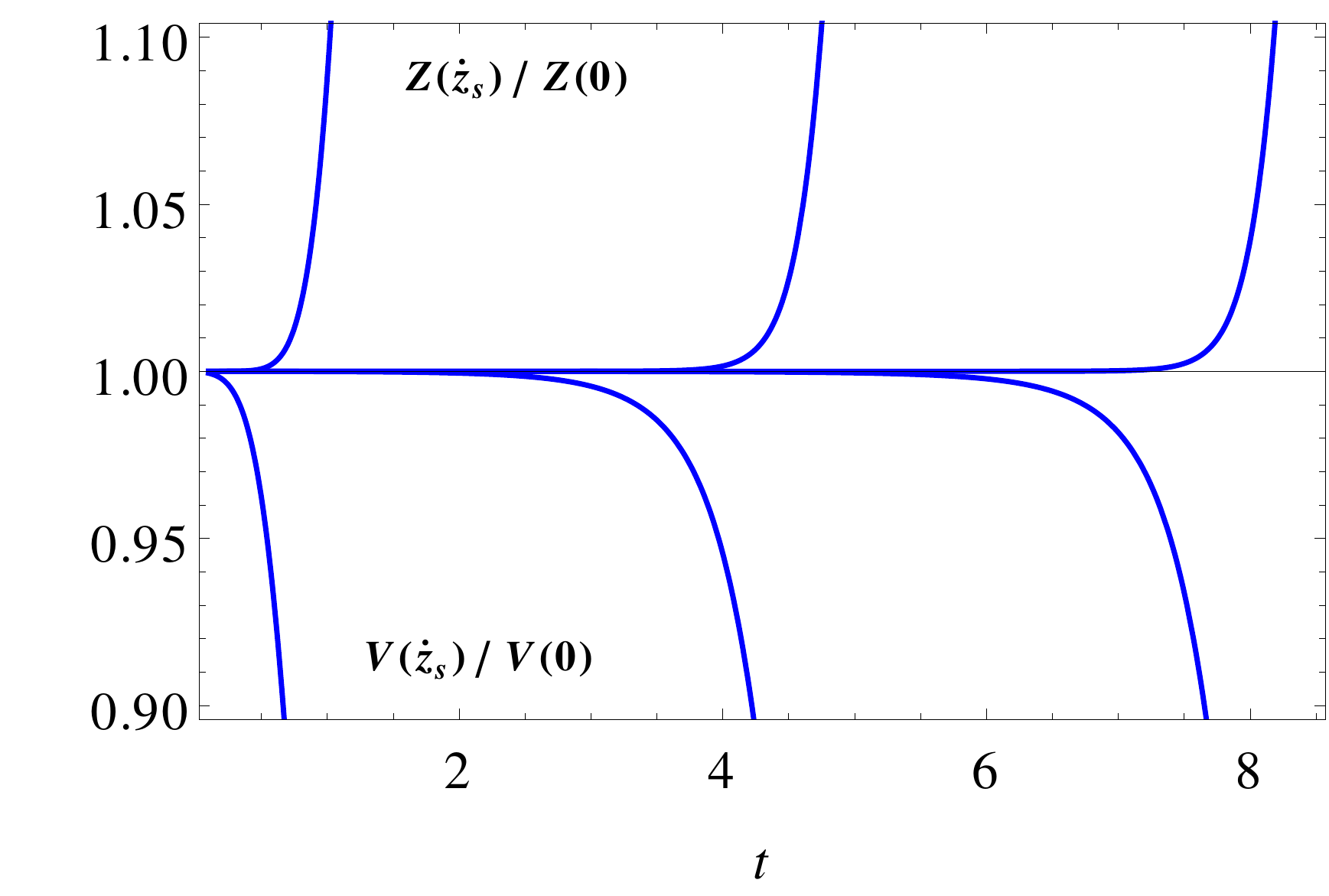}
\caption{\label{fig:relmc} 
A plot of the ratios $Z(\dot{z}_s)/Z(0)$ (upper part) $V(\dot{z}_s)/V(0)$ (lower part) as functions of the boundary time. In both cases, the three curves correspond to the three cases displayed in fig.~\ref{fig:staticapprox}, i.e.~$z_0=0.2$ and $c=0,\;0.33,\;1/3$ (from left to right). 
 }
\end{center}
\end{figure}

To understand the observed behavior, it is instructive to inspect, how the full matching conditions of eq.~(\ref{eq:matching}) relate to the static approximation of eq.~(\ref{qs}). The result of this comparison is displayed in fig.~\ref{fig:relmc}, where we plot $Z(\dot{z}_s)/Z(0)$ in the upper and $V(\dot{z}_s)/V(0)$ in the lower part of the figure, both given as functions of the boundary time. Shown are three curves corresponding to the three different values of the shell initial data already inspected in fig.~\ref{fig:staticapprox}: $c=0$, 0.3 and 1/3, with $z_0$ set to 0.2 in each case. Comparing to fig.~\ref{fig:staticapprox}, we observe that in all three cases the deviation of the quasistatic HEE from the full result can to a good accuracy be attributed to the growth of the derivative terms in the junction conditions. In particular, the onset of the rapid growth of $Z(\dot{z}_s)/Z(0)$ in fig.~\ref{fig:relmc} coincides very accurately with the point of time, when the quasistatic entropy starts to visibly deviate from the full result in fig.~\ref{fig:staticapprox}. 

Having seen that the success of the quasistatic approximation can be traced back to the junction conditions for extremal surfaces, it is natural to ask, to what extent one can understand the reason for the slow turning on of their derivative terms. To this end, we now switch to the $r$, $t$ coordinate system, in which the matching conditions can be shown to take the forms
\begin{align}
\frac{t'_+}{r'_-}\equiv T(\dot{r}_s)=\frac{\dot{r}_s}{f_+ f_-}(\beta_--\beta_+)\,,
\\
\frac{r'_+}{r'_-}\equiv R(\dot{r}_s)=\frac{1}{f_-}(\beta_+\beta_--\dot{r}_s^2)\,,
\end{align}
with the quasistatic limit corresponding to 
\beq
T(0)=0,\quad R(0)=\sqrt{\frac{f_+}{f_-}}\;,
\eeq
and the Vaidya one to
\beq
T(\infty)=-\frac{f_+-f_-}{2f_+f_-},\quad R(\infty)=\frac{f_++f_-}{2f_-}\,.
\eeq
Expanding these functions around $r=\infty$ in the case of a $d$-dimensional field theory, we obtain in the quasistatic approximation
\beq
T(0)=0\quad R(0)=1-\frac{1}{2r^d}+O(r^{-2d}),
\eeq
while the Vaidya counterparts of these results become
\beq
T(\infty)=\frac{1}{2 r^{d+2}},\quad R(\infty)=1-\frac{1}{2r^d}\,.
\eeq
From here, we see that for $r\gg 1$ the two matching conditions quickly approach each other, and that in particular the asymptotic behavior of $R$ is independent of $\dot{r}_s$. This explains, why the quasistatic entropy approximates the full results, and even the Vaidya limit, so well, when the shell is released from close to the boundary.

\begin{figure}[t]
\includegraphics[scale=0.67]{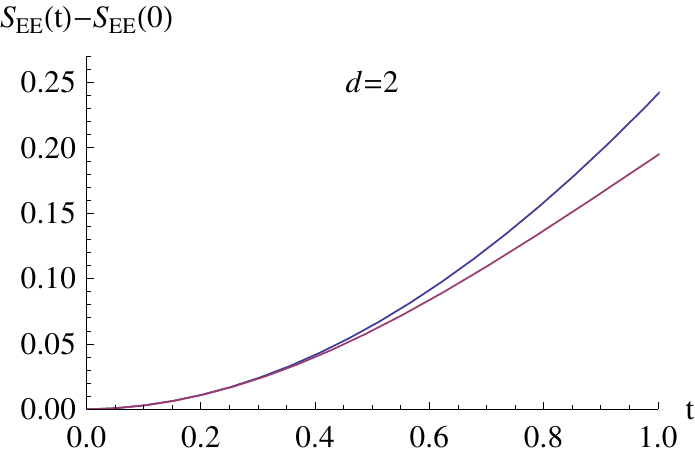}$\;\; $\includegraphics[scale=0.67]{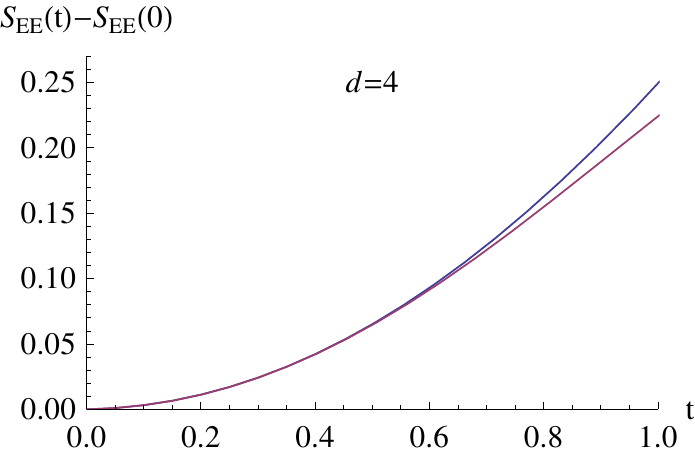}$\;\; $\includegraphics[scale=0.67]{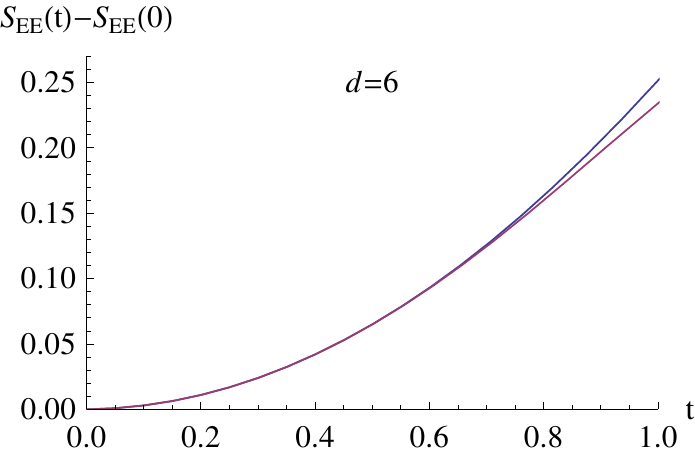}
\caption{\label{fig:quasivaidya2} Comparison of the Vaidya (blue curve) and quasistatic (purple curve) limits of the HEE at $d=2$, 4 and 6, evaluated for a strip with $L=8$.
}
\end{figure}

Finally, we remark that it is clear from the above results that the quasistatic approximation should successively improve, as the number of spatial dimensions in the system is increased --- a direct consequence of the $d$-dependence of the blackening factor in the AdS Schwarzschild metric, $h(z)=1-z^d$. This effect is indeed clearly seen in the three plots displayed in fig.~\ref{fig:quasivaidya2}, where we compare the quasistatic and Vaidya limits of the HEE at $d=2$, 4 and 6. In the quasistatic results, the trajectory of the shell is taken to be lightlike and start from the boundary.

%
%
%

\section{Conclusions}

Many of the existing studies of holographic equilibration have been performed within highly simplified models, which however have the virtue of allowing the determination of rather complicated physical quantities. One prominent example of this is the description of black hole formation via the gravitational collapse of a thin shell of matter in AdS spacetime, typically motivated as resulting from a rapid quench in the dual field theory system. Up to very recently, the determination of physical observables in this model has, however, been only possible with further simplifications, such as approximating the spacetime by its Vaidya limit, corresponding to a lightlike shell, or in the quasistatic approximation where the shell is taken to move arbitrarily slowly when formulating the so-called junction conditions. In our previous paper \cite{Keranen:2014zoa}, we took the first steps towards overcoming these limitations in the case of the Holographic Entanglement Entropy (HEE). In particular, we considered there the time evolution of the HEE in the background of a shell following its physical trajectory, varying both the equation of state and turning point of the shell. Doing so, we were able to verify the earlier conjecture of \cite{Liu:2013iza} concerning the existence of a linear regime in the time evolution of the HEE that only depends on the properties of the final state of the system.

In the paper at hand, we have continued work in the direction of \cite{Keranen:2014zoa}. In particular, we have studied the universality of the early and late time behaviors of the evolution of the HEE, and provided the first ever comparison of a fully dynamical thermalization calculation with its Vaidya and quasistatic limits. As expected, we observed that the quasistatic results provide a good approximation of the full ones at early times, while the Vaidya limit is approached at late times. Similarly, the quality of the quasistatic approximation improves when the shell EoS is chosen to minimize the rate of the collapse, while the Vaidya limit works better when the shell is allowed to accelerate more quickly. What we, however, found surprising was the fact that for many shell EoSs one can identify a brief period of overlap between the regions of validity of the quasistatic and Vaidya approximations, which seems to be at odds with the opposite nature of the two limits. To this end, we performed a detailed investigation of the reason of this behavior, tracing it back to the form of the junction conditions for extremal surfaces at the shell and to the slow onset of derivative terms in them. This observation suggests that the behavior may well be of somewhat universal nature, extending beyond the HEE, and raises hopes that the finding may eventually be used to simplify the holographic determination of many other dynamical quantities.

Apart from improving the general understanding of thermalization dynamics, the most important outcome of our present work is the introduction of new technical tools to aid dynamical calculations within the collapsing shell model. In particular, the construction of a coordinate system continuous at the location of the shell and the derivation of junction conditions for extremal surfaces penetrating the shell are results that we hope will find applications in many forthcoming works.


\section*{Acknowlegements}
We thank Rudolf Baier, Jan de Boer, Niko Jokela, Esko Keski-Vakkuri, and Javier Mas for useful discussions. V.K.~was supported by the European Research Council under the European Union's Seventh Framework Programme (ERC Grant agreement 307955), H.N.~by the Bielefeld Young Researcher's Fund, S.S.~by the FWF project P26328, O.T.~by a research program of the "Stichting voor Fundamenteel Onderzoek der Materie (FOM)", financially supported by the "Nederlandse organisatie voor Wetenschappelijke Onderzoek (NWO)", and~A.V. by the Academy of Finland, grant $\#$266185.

\begin{appendix}
 
\section{Equation of motion for the shell\label{sec:eom}}
In this first appendix, we derive the equation of motion of a thin shell undergoing gravitational collapse in AdS$_5$ spacetime. To simplify our expressions, we suppress here the indices $\pm$ indicating whether we are inside or outside of the shell. All identities involving $f$ or $t$ can be seen to hold both inside and outside of the shell.

The unit normal vector of the shell is easily seen to read
\begin{equation}
n^r = \sqrt{f+\dot{r}_s^2} \quad \text{and} \quad n^t = \frac{\dot{r}_s}{f} \, ,
\end{equation}
where we have chosen the vector to point towards increasing $r$, and $f$ is evaluated at the location of the shell, $r=r_s$. Using this, we see that the only nonzero components of the shell's extrinsic curvature, 
\begin{equation}
K_{ij} = n_\alpha\left( \frac{\partial^2 y^\alpha}{\partial \xi^i \, \partial \xi^j} + \Gamma^\alpha_{\beta\gamma} \frac{\partial y^\beta}{\partial \xi^i}\frac{\partial y^\gamma}{\partial \xi^j}\right) \, ,
\end{equation}
are
\begin{equation}
K_{\tau\tau} = \frac{f'+2\ddot{r}_s}{2\sqrt{f+\dot{r}_s^2}} \quad \text{and} \quad K_{xx} = K_{yy} = K_{zz} = -r_s\sqrt{f+\dot{r}_s^2} \; .
\end{equation}
The induced metric on the shell is on the other hand given by
\begin{equation}
\gamma_{ij} = \partial_i y^\mu\,\partial_j y^\nu\,g_{\mu\nu}\, ,
\end{equation}
so that
\begin{equation}
\gamma_{\tau\tau} = -1 \quad \text{and} \quad \gamma_{xx} = \gamma_{yy} = \gamma_{zz} = r^2_s \, .
\end{equation}
Thus, the trace of the curvature tensor reads
\begin{equation}
K  = \gamma^{ij} K_{ij} = -\frac{1}{\sqrt{f+\dot{r}_s^2}}\left( \ddot{r}_s+\frac{f'}{2}+\frac{3}{r_s}\left( f+\dot{r}_s^2\right)\right)\, .
\end{equation}

To derive the EoM of the shell, we use the Isreal junction condition \cite{Israel:1966rt}. It states that the difference of the extrinsic curvature between the inside and outside of the shell is related to the energy-momentum content of the object through
\begin{equation}
\label{eq:Israeli} \left[ K_{ij}-\gamma_{ij}K\right] = -8\pi g_5 S_{ij} \, ,
\end{equation}
where the square brackets denote the difference between the inside and outside,
\begin{equation}
\left[ \mathcal{O} \right] = \mathcal{O}_\mathrm{inside} - \mathcal{O}_\mathrm{outside}\, .
\end{equation}
The LHS of eq.~(\ref{eq:Israeli}) is zero for the non-diagonal terms, while for the diagonal terms the expression inside the square bracket is given by
\begin{align}
K_{\tau\tau} - \gamma_{\tau\tau} K & = -\frac{3}{r_s} \sqrt{f+\dot{r}_s^2} \, , \\
K_{xx}-\gamma_{xx} K & = \frac{1}{\dot{r}_s} \frac{d}{d\tau}\left( r_s^2 \sqrt{f+\dot{r}_s^2}\right) \, .
\end{align}

\section{Determining the metric in the continuous coordinate system \label{sec:cont}}

In this appendix, we determine the values of the partial derivatives $\PD{r,\tau,\lambda}$, $\PD{t,\tau,\lambda}$, etc.~needed to construct a coordinate system continuous at the shell, cf.~section \ref{sec:junctionconditions}.  

The equations determining spacelike geodesics are given by
 \begin{align}
\label{eq:geodesic1} \frac{f'}{f} \frac{dr}{d\sigma}\frac{dt}{d\sigma} + \frac{d^2t}{d\sigma^2}  & \;=\; 0\,,\\
\label{eq:geodesic2}-\frac{f'}{2f}\left( \frac{dr}{d\sigma} \right)^2 +\frac{1}{2} f\,f'\left( \frac{dt}{d\sigma} \right)^2 - r\,f\left( \frac{dx}{d\sigma} \right)^2 + \frac{d^2r}{d\sigma^2}  &\;=\;0\,,\\
\label{eq:geodesic3}\frac{2}{r}\frac{dr}{d\sigma}\frac{dx}{d\sigma} + \frac{d^2x}{d\sigma^2}  &\;=\;0\,,
\end{align}
where $'$ denotes a derivative with respect to $r$, and $\sigma$ is the affine parameter of the geodesic, identified as the proper length $\sigma = \lambda$. The trajectories we are interested in are defined at a constant $\mathbf{x}$, so that the equations can be integrated to give 
\begin{align}
\label{eq:ftdot} f\frac{dt}{d\lambda} & = A\, ,\\
\label{eq:rdotsquared} \left(\frac{dr}{d\lambda}\right)^2 & = f + A^2 \, .
\end{align}
The value of the integration constant $A$ can be fixed by requiring that the geodesic points in the direction of the normal vector of the shell at its location,
\begin{equation}
\left.\label{eq:normal} \frac{dx^\mu}{d\lambda} \right|_{\lambda=0}= n^\mu \, ,
\end{equation}
where $n^\mu$ has the form
\begin{equation}
\left[ n^\mu \right] = \left( \dot{r}_s/f, \sqrt{f+\dot{r}_s^2},\vec{0}\right) \, .
\end{equation}
Evaluating eq.~(\ref{eq:normal}) at the shell, we easily get
\begin{align}
\frac{dt}{d\lambda} &= \frac{A}{f(r_s)} = \frac{\dot{r}_s}{f(r_s)}\,,\\
\frac{dr}{d\lambda} &= \sqrt{f(r_s) + A^2} = \sqrt{f(r_s) + \dot{r}_s^2} \, ,
\end{align}
from which we see that $A = \dot{r}_s$.

Since eqs.~(\ref{eq:ftdot}) - (\ref{eq:rdotsquared}) apply for spacelike geodesics, which by definition have constant $\tau$, these relations immediately produce the two partial derivatives
\begin{equation}
\PD{t,\lambda,\tau} = \frac{\dot{r}_s}{f(r)} \quad \text{and} \quad \PD{r,\lambda,\tau} = \sqrt{f(r) + \dot{r}_s^2} \; .
\end{equation}
To get the derivatives $\PD{r,\tau,\lambda}$ and $\PD{t,\tau,\lambda}$, we on the other hand need to differentiate the integrals of the equations of motion. Equation (\ref{eq:rdotsquared}) can be integrated to yield
\begin{equation}
\label{eq:lambdar} \int_0^\lambda d\lambda = \int_{r_s(\tau)}^{r(\tau,\lambda)} \frac{dr}{\sqrt{f(r)+\dot{r}^2_s(\tau)}} \,,
\end{equation}
which upon differentiation w.r.t.~$\tau$ leads to the expression
\begin{equation}
\PD{\lambda,\tau,\lambda} =  \PD{r,\tau,\lambda} \frac{1}{\sqrt{f+\dot{r}_s^2}} -\frac{\dot{r}_s}{\sqrt{f_s+\dot{r}_s^2(\tau)}} - \int_{r_s(\tau)}^{r(\lambda,\tau)}  dr \frac{\ddot{r}_s\dot{r}_s}{(f(r) + \dot{r}_s^2)^{\frac{3}{2}}} \, .
\end{equation}
By changing the integration variable in this equation to $\lambda$ and requiring that the expression vanishes ($\lambda$ and $\tau$ are by definition independent variables) then gives
\begin{equation}
\PD{r,\tau,\lambda} = \dot{r}_s\sqrt{f+\dot{r}_s^2}\left(\frac{1}{\sqrt{f_s+\dot{r}_s^2}} +  \int_0^\lambda d\lambda \frac{\ddot{r}_s}{f+\dot{r}_s^2} \right) \label{eq:drds} \, .
\end{equation}
Finally, to get $\PD{t,\tau,\lambda}$, we solve $t$ as an integral of $\lambda$ from eq.~(\ref{eq:ftdot}) and then differentiate it w.r.t.~$\tau$ to get
\begin{equation}
\label{eq:dtds1} \PD{t,\tau,\lambda} - \dot{t}_s = \int_0^\lambda d\lambda \left[ \frac{\ddot{r}_s}{f} - \dot{r}_s\frac{f'}{f^2}\PD{r,\tau,\lambda}\, \right] \, ,
\end{equation}
which quickly leads us to 
\begin{equation}
\label{eq:dtds}
\PD{t,\tau,\lambda} = \frac{f+\dot{r}_s^2}{f}\left(\frac{1}{\sqrt{f_s+\dot{r}_s^2}} +  \int_0^\lambda d\lambda \frac{\ddot{r}_s}{f+\dot{r}_s^2} \right) = \frac{\sqrt{f+\dot{r}_s^2}}{f \, \dot{r}_s} \PD{r,\tau,\lambda} \, .
\end{equation}

Now that we have all the necessary partial derivates at hand, we can see what our desired metric looks like. Setting $dx=0$, we get
\begin{align*}
ds^2 & = -f\,dt^2 + \frac{dr^2}{f}\\
& = d\tau^2 \left[-f \PD{t,\tau,\lambda}^2 + \frac{1}{f} \PD{r,\tau,\lambda}^2\right] + d\lambda^2 \left[ -f \PD{t,\lambda,\tau}^2+ \frac{1}{f} \PD{r,\lambda,\tau}^2\right]\\
& \qquad + 2\,d\lambda\,d\tau\,\left[ -f\PD{t,\tau,\lambda}\PD{t,\lambda,\tau}+\frac{1}{f}\PD{r,\tau,\lambda}\PD{r,\lambda,\tau} \right]\,,
\end{align*}
where the $\lambda\lambda$-component can be simplified using the fact that $\lambda$ was defined as a proper length,
\begin{equation}
g_{\lambda\lambda} = -f \PD{t,\lambda,\tau}^2 + \frac{1}{f} \PD{r,\lambda,\tau}^2 = -f \frac{\dot{r}_s^2}{f^2}+ \frac{1}{f}(f+\dot{r}_s^2) = 1 \, .
\end{equation}
The non-diagonal part of the metric is on the other hand given by
\begin{align}
\nonumber g_{\tau\lambda} = & -f \PD{t,\tau,\lambda}\PD{t,\lambda,\tau} + \frac{1}{f} \PD{r,\tau,\lambda}\PD{r,\lambda,\tau}\\
= & -f \frac{\dot{r}_s}{f} \frac{\sqrt{f+\dot{r}_s^2}}{f \, \dot{r}_s} \PD{r,\tau,\lambda} + \frac{1}{f} \sqrt{f+\dot{r}_s^2} \PD{r,\tau,\lambda}\\
\nonumber = & \, 0 \, ,
\end{align}
implying that the metric is everywhere diagonal. Finally, the $\tau\tau$-component is given by the (by construction continuous) function
\begin{align}
\nonumber g_{\tau\tau} = & -f \PD{t,\tau,\lambda}^2 + \frac{1}{f} \PD{r,\tau,\lambda}^2\\
= & -f \frac{f+\dot{r}_s^2}{f^2 \dot{r}_s^2}\PD{r,\tau,\lambda}^2 + \frac{1}{f} \PD{r,\tau,\lambda}^2\\
\nonumber = & -\left(\sqrt{\frac{f+\dot{r}_s^2}{f_s+\dot{r}_s^2}} + \sqrt{f+\dot{r}_s^2} \int_0^\lambda d\lambda \frac{\ddot{r}_s}{f+\dot{r}_s^2} \right)^2 \, .
\end{align}
It is important to note that although the result for the $\tau\tau$ component of the metric looks complicated and has dependence on the function $f$, on the shell it is equal to just $-1$, independent of the functional form of $f(r)$. This reflects the fact that $\tau$ was defined as the proper time of the shell.

\section{Generalized junction conditions \label{genjun}}

In this appendix, our goal is to generalize the junction conditions to the case where there are two unknown functions in the metric,
\begin{equation}
ds^2 = -f(r) \,dt^2 + \frac{dr^2}{g(r)} + r^2\,d\mathbf{x}^2 \, .
\end{equation}
Following the above treatment, we again first construct the continuous coordinate system, and then proceed to derive the junction conditions.

We will again take the location of the shell to be parameterized by $(t_s(\tau),r_s(\tau))$; this time the relation between $r_s$ and $t_s$, however, reads
\begin{equation}
\dot{t}_s^2 = \frac{1}{f}\left( \frac{\dot{r}_s^2}{g}+1\right) \, ,
\end{equation}
while the normal vector of the shell is given by
\begin{equation}
\left[ n^\mu \right] = \left( \frac{\dot{r}_s}{\sqrt{fg}}, \sqrt{g+\dot{r}_s^2},0,\ldots\right) \, .
\end{equation}
The geodesic equations are now seen to take the forms
\begin{align}
\frac{f'\dot{r}\dot{t}}{f}+\ddot{t} & = 0,\,\\
-\frac{g'}{2g}\dot{r}^2 + \frac{1}{2} g f' \dot{t}^2 + \ddot{r} & =0,\,
\end{align}
which --- taking into account that the affine parameter is the proper length of a space-like geodesic --- can be integrated to
\begin{align}
f\,\dot{t} & =A\, ,\\
\dot{r}^2 &= g\left( 1+\frac{A^2}{f}\right)\, .
\end{align}
Requiring finally that the geodesic is normal to the shell,
\begin{equation}
\frac{dx^\mu}{d\lambda}\bigg|_{\lambda=0} = n^\mu \, ,
\end{equation}
we obtain for the constant $A$
\begin{equation}
A = \sqrt{\frac{f_s}{g_s}} \dot{r}_s\, .
\end{equation}

At this point, we can again read off the necessary partial derivatives at the shell, obtaining 
\begin{equation}
\PD{t,\lambda,\tau} = \frac{\dot{r}_s}{\sqrt{f_sg_s}}\;, \quad \PD{r,\lambda,\tau} = \sqrt{g_s+\dot{r}_s^2}\;,\quad \PD{t,\tau,\lambda} = \sqrt{\frac{g_s+\dot{r}_s^2}{f_s g_s}}\;,\quad\PD{r,\tau,\lambda}=\dot{r}_s \, .
\end{equation}
Inserting these into the relations (cf.~eqs.~(\ref{eq:dlambdadx})--(\ref{eq:dsdx}))
\begin{align}
\frac{d\lambda}{d\sigma} & = \frac{\PD{r_+,\tau,\lambda}\frac{dt_+}{d\sigma}-\PD{t_+,\tau,\lambda}\frac{dr_+}{d\sigma}}{\PD{r_+,\tau,\lambda}\PD{t_+,\lambda,\tau}-\PD{r_+,\lambda,\tau}\PD{t_+,\tau,\lambda}}\,,\\
\frac{d\tau}{d\sigma} & = \frac{\PD{r_+,\lambda,\tau}\frac{dt_+}{d\sigma}-\PD{t_+,\lambda,\tau}\frac{dr_+}{d\sigma}}{\PD{r_+,\lambda,\tau}\PD{t_+,\tau,\lambda}-\PD{r_+,\tau,\lambda}\PD{t_+,\lambda,\tau}} \, ,
\end{align}
we get at the location of the shell
\begin{align}
\frac{d\lambda}{d\sigma} &= -\sqrt{\frac{f_+}{g_+}}\dot{r}_s \frac{dt_+}{d\sigma}+\frac{\sqrt{g_++\dot{r}_s^2}}{g_+}\frac{dr_+}{d\sigma}\,, \\
\frac{d \tau}{d\sigma} & = \sqrt{\frac{f_+}{g_+}}\sqrt{g_++\dot{r}_s^2}\frac{dt_+}{d\sigma} -\frac{\dot{r}_s}{g_+}\frac{dr_+}{d\sigma} \, .
\end{align}
Using the chain rule, we next express $dt/d\sigma$ and $dr/d\sigma$ inside of the shell as
\begin{align}
\frac{dt_-}{d\sigma} &= \PD{t_-,\tau,\lambda}\frac{d\tau}{d\sigma} + \PD{t_-,\lambda,\tau}\frac{d\lambda}{d\sigma}\,,\\
\frac{dr_-}{d\sigma} &= \PD{r_-,\tau,\lambda}\frac{d\tau}{d\sigma} + \PD{r_-,\lambda,\tau}\frac{d\lambda}{d\sigma}\,,
\end{align}
which, when evaluated at the shell, produce
\begin{align}
\frac{dt_-}{d\sigma} &= \sqrt{\frac{g_-+\dot{r}_s^2}{f_-g_-}}\frac{d\tau}{d\sigma}+\frac{\dot{r}_s}{\sqrt{f_-g_-}}\frac{d\lambda}{d\sigma}\,,\\
\frac{dr_-}{d\sigma} &= \dot{r}_s \frac{d\tau}{d\sigma} +\sqrt{g+\dot{r}_s^2}\frac{d\lambda}{d\sigma} \, .
\end{align}
Combining finally all the above results, we obtain as the generalized junction conditions
\begin{align}
\frac{dt_-}{d\sigma} & = \frac{dt_+}{d\sigma}\sqrt{\frac{f_+}{f_-g_-g_+}}\left( \beta_+\beta_- -\dot{r}_s^2\right)+\frac{dr_+}{d\sigma} \frac{\dot{r}_s}{g_+\sqrt{f_-g_-}}\left( \beta_+ - \beta_-\right)\, ,\\
\frac{dr_-}{d\sigma} & = \frac{dt_+}{d\sigma} \dot{r}_s\sqrt{\frac{f_+}{g_+}}\left(\beta_+-\beta_-\right)+\frac{dr_+}{d\sigma} \frac{1}{g_+}\left( \beta_+\beta_--\dot{r}_s^2\right)\,,
\end{align}
where $\beta_\pm$ are defined as before and the expression is evaluated at the location of the shell.

\section{Spherical boundary region}\label{sec:spherical}
In this last appendix, we provide some details for the computation of the HEE in the case, where the boundary surface has the form of a sphere of radius $R$. In this case, the extremal surface is independent of the angular coordinates due to rotational symmetry, and we can parametrize it as $z=z(\rho)$, $v=v(\rho)$, with $\rho$ being the radial coordinate on the field theory side (i.e.~on the boundary). The area functional becomes then
\beq
A=4\pi \int d\rho\frac{\rho^2}{z(\rho)^3}\sqrt{B}=4\pi \int d\rho \, \mathcal{L}\,,
\eeq
where $B$ is the same quantity as in the strip case except that the derivatives $z'$ and $v'$ are derivatives with respect to $\rho$. 

Due to the explicit appearance of $\rho$ in the area functional, there are fewer conserved quantities this time, making the spherical case slightly more complicated than the strip one. There is, however, still a partial time translational invariance away from the shell, which gives rise to the conservation law
\beq
\frac{\partial\mathcal{L}}{\partial v'}=-\frac{\rho^2(h v'+z')}{z^3\sqrt{B}}=E\,,\label{eq:energy2}
\eeq
with $E$ again taking different values on the two sides of the shell. In the interior region (assuming the boundary radius to be large enough so that the extremal surface passes through the shell) there is a turning point at $\rho=0$, where $z'$ and $v'$ vanish, which immediately tells us that $E=0$ in the interior, or
\beq
v'=-z'\,.
\eeq
Applying this identity in the Euler-Lagrange equation for $z(\rho)$, we obtain
\beq
z(\rho z''+2 z'^3+2 z')+3 \rho (1+z'^2)=0\,.
\eeq
This equation has a one parameter family of solutions
\beq
z(\rho)=\sqrt{z_*^2-\rho^2}\,,
\eeq
labeled by the turning point $z_*$, identified as the most general regular solution with a turning point at $z=z_*$. For a second order equation, one should specify two initial conditions, which in our case are chosen as $z(0)=z_*$ and $z'(0)=0$.

Outside the shell, we use eq.~(\ref{eq:energy2}) to solve for $v'$ and plug this into the Euler-Lagrange equation for $z$, giving
\beq
2(\rho^4 z h+E_+^2z^7)z''+2\rho^3 h (3\rho z'^2+2zz')+z(\rho^3 z'^2(-\rho h'+4 z')+E_+^2z^6 h')+6\rho^4 h^2=0\,.
\eeq
This equation needs to be solved numerically. Noting that the interior surface satisfies $v'=-z'$, the junction condition for the derivatives at the position of the shell is again given by eq.~(\ref{eq:matching}), with the derivatives now understood as derivatives with respect to $\rho$. This way the value of the constant $E_+$
is fixed to be
\beq
E_+=-\frac{\rho_c^2(h_+ v'_++z'_+)}{z_c^3\sqrt{1-h_+ (v_+')^2-2v_+'z_+'}}\,,\label{eq:energy3}
\eeq
which completes our exercise.

\end{appendix}

\bibliographystyle{JHEP-2} 

\bibliography{longpaper1}

\end{document}